\documentclass[twocolumn,aps,prc,superscriptaddress]{revtex4-2}
\usepackage{color,amsmath,amssymb,epsfig,graphicx}
\usepackage{bm}
\usepackage[colorlinks=true,linkcolor=blue,filecolor=blue,urlcolor=blue,citecolor=blue]{hyperref}
\usepackage{mathptmx, courier, pifont}
\usepackage[scaled=0.92]{helvet}
\usepackage[T1]{fontenc}
\usepackage{textcomp}
\usepackage{multirow}
\usepackage{longtable}

\begin{document}

\title{ Microscopic investigation of $E2$ matrix elements in atomic nuclei - II }

\author{Kouser Qureshie} 
\affiliation{Department of  Physics, Islamic University of Science and Technology, Awantipora, 192 122, India}
\author{S. P. Rouoof} \email{sprouoofphysics27@gmail.com}
\affiliation{Department of  Physics, Islamic University of Science and Technology, Awantipora, 192 122, India}
\author{J. A. Sheikh} \email{sjaphysics@gmail.com}
\affiliation{Department of Physics, University of Kashmir, Srinagar, 190 006, India}
\author{N. Rather}
\affiliation{Department of  Physics, Islamic University of Science and Technology, Awantipora, 192 122, India}
\author{S. Jehangir}
\affiliation{Department of Physics, Government Degree College Kulgam, Jammu and Kashmir, 192 231, India}
\author{G. H. Bhat}
\affiliation{Department of Physics, Government Degree College Shopian, Jammu and Kashmir, 192 303, India}
\author{ S. Frauendorf}
\affiliation{Department of Physics, University of Notre Dame, Notre Dame, Indiana 46556,  USA}

\date{\today}

\begin{abstract}

  The present work is a continuation of our earlier investigation with the primary objective to systematically
  calculate the $E2$ matrix elements using the microscopic approach of the triaxial projected shell model (TPSM).
  In the earlier work, we studied nine nuclides of $^{72}$Ge, $^{76}$Ge, $^{104}$Ru, $^{168}$Er, $^{186}$Os, $^{188}$Os, $^{190}$Os, $^{192}$Os, and $^{194}$Pt. In
  the present work six more nuclides of $^{70}$Ge, $^{76,78,80,82}$Se, and $^{100}$Mo have been investigated.
  The Coulomb excitation data has
  recently become available for $^{70}$Ge and other nuclides were inadvertently
  omitted in our earlier investigation. It is demonstrated that TPSM approach provides a good description of the
  available experimental data and most of the nuclides, except for $^{76}$Se and $^{100}$Mo, are shown to have
  $\gamma$ soft behaviour. Further, it is
  demonstrated that in contrast to the predictions of the phenomenological collective model, TPSM calculations depict no
  clear correlation
  between the energy staggering pattern of the $\gamma$ band and the deduced shape invariant quantities
  using the Kumar-Cline sum rules.
   
\end{abstract}


\maketitle

\section{INTRODUCTION}\label{Intro}

To elucidate the shapes of atomic nuclei continues to be one of the challenging problems in nuclear physics \cite{BMII,RS80}. The shape of a nucleus is
not only important in nuclear physics, but also plays a crucial role in other areas of physics. For instance, the pear-shaped nuclides, $^{220}$Rn
and $^{224}$Ra, recently studied at ISOLDE (CERN) \cite{gaffney}, may provide answers on the fundamental question of matter-antimatter imbalance
in the universe. The shape of a nucleus is ascertained through electromagnetic probes, and as a matter of fact
the large quadrupole moments observed for certain nuclei in the early development of nuclear physics provided a first indication that nuclei
could be deformed \cite{rainwater,ab52,BMII}. There are primarily two experimental methods to determine the electromagnetic transition probabilities, one is
through lifetime measurements and the other is using the Coulomb excitation (COULEX) mechanism \cite{DC86}. The COULEX experiment has the advantage over the
lifetime measurement since it can determine the sign of the quadrupole matrix elements. It is, therefore, possible to resolve whether an
axial nucleus has a prolate or an oblate shape using the COULEX approach \cite{DC86}.

In the early days of nuclear physics, it was possible to populate only low-spin states with light-ion beam used in the
COULEX experiments. However, the advancements in the accelerator technology and detection techniques has made it possible to
perform COULEX experiments with heavier beams, and populate nuclei up to an excitation energy of 2 MeV. The major advantage of the
COULEX technique is that a complete set of $E2$ matrix elements can be deduced by performing the analysis of the data using the advanced
computer codes, such as, GOSIA \cite{GOSIA2012}. A complete set of $E2$ matrix elements then allows to determine the nuclear shape
in a model independent manner using the Kumar-Cline sum rules \cite{DC86,KM72}. In this approach, the $E2$ spherical tensor
operators are expressed in terms of rotationally invariant quantities. These invariant quantities are then related to the ellipsoidal shape
parameters, which are similar to $\beta$ and $\gamma$ parameters of the Bohr-Mottelson collective model \cite{DC86,KM72,GOSIA2012}.

\begin{figure}[!h]
 \centerline{\includegraphics[trim=0cm 0cm 0cm
0cm,width=0.5\textwidth,clip]{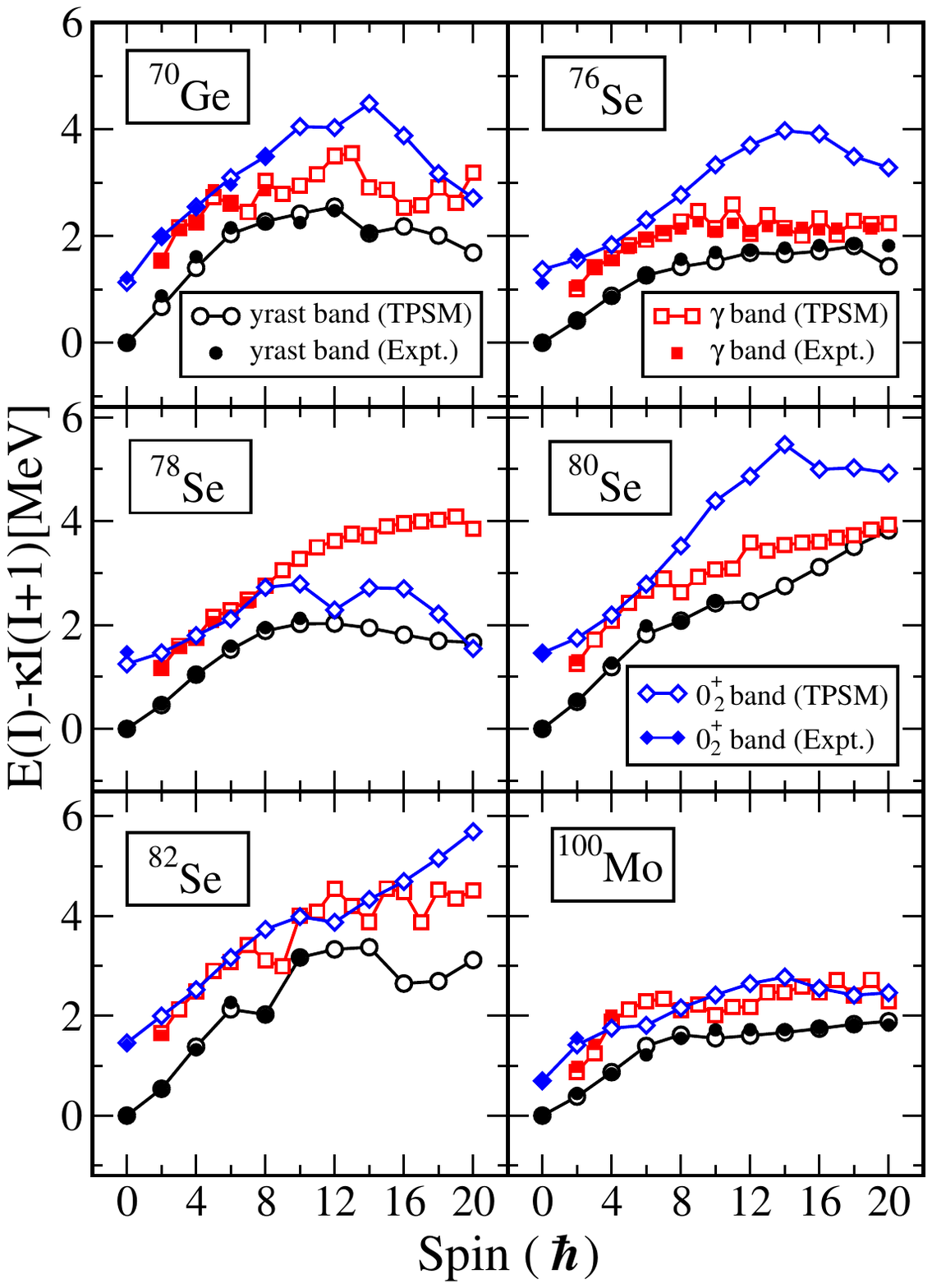}} \caption{(Color
   online) TPSM and experimental energies of the yrast, $\gamma$, and $0_2^+$ bands of $^{70}$Ge, $^{76,78,80,82}$Se, and $^{100}$Mo isotopes. The scaling factor,
   $\kappa=32.32A^{-5/3}$.
  }
\label{E_core}
\end{figure}

\begin{figure}[!h]
 \centerline{\includegraphics[trim=0cm 0cm 0cm
0cm,width=0.5\textwidth,clip]{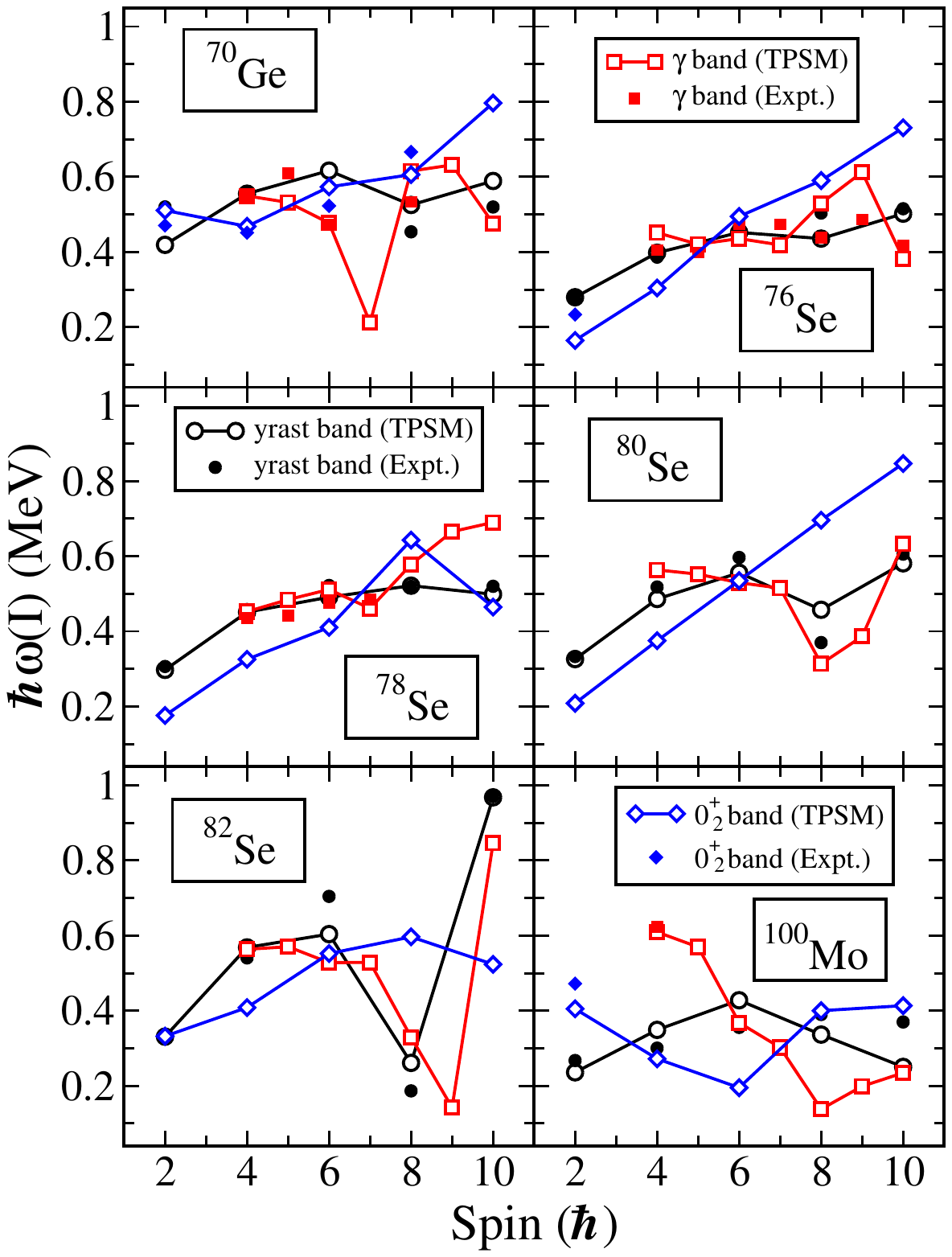}} \caption{(Color
online) Frequency $\hbar \omega(I)=(E(I)-E(I-2))/2$ for  the yrast, $\gamma$, and $0_2^+$ bands of $^{70}$Ge, $^{76,78,80,82}$Se,and $^{100}$Mo isotopes. 
  }
\label{E_core}
\end{figure}

\begin{figure}[!h]
 \centerline{\includegraphics[trim=0cm 0cm 0cm
0cm,width=0.5\textwidth,clip]{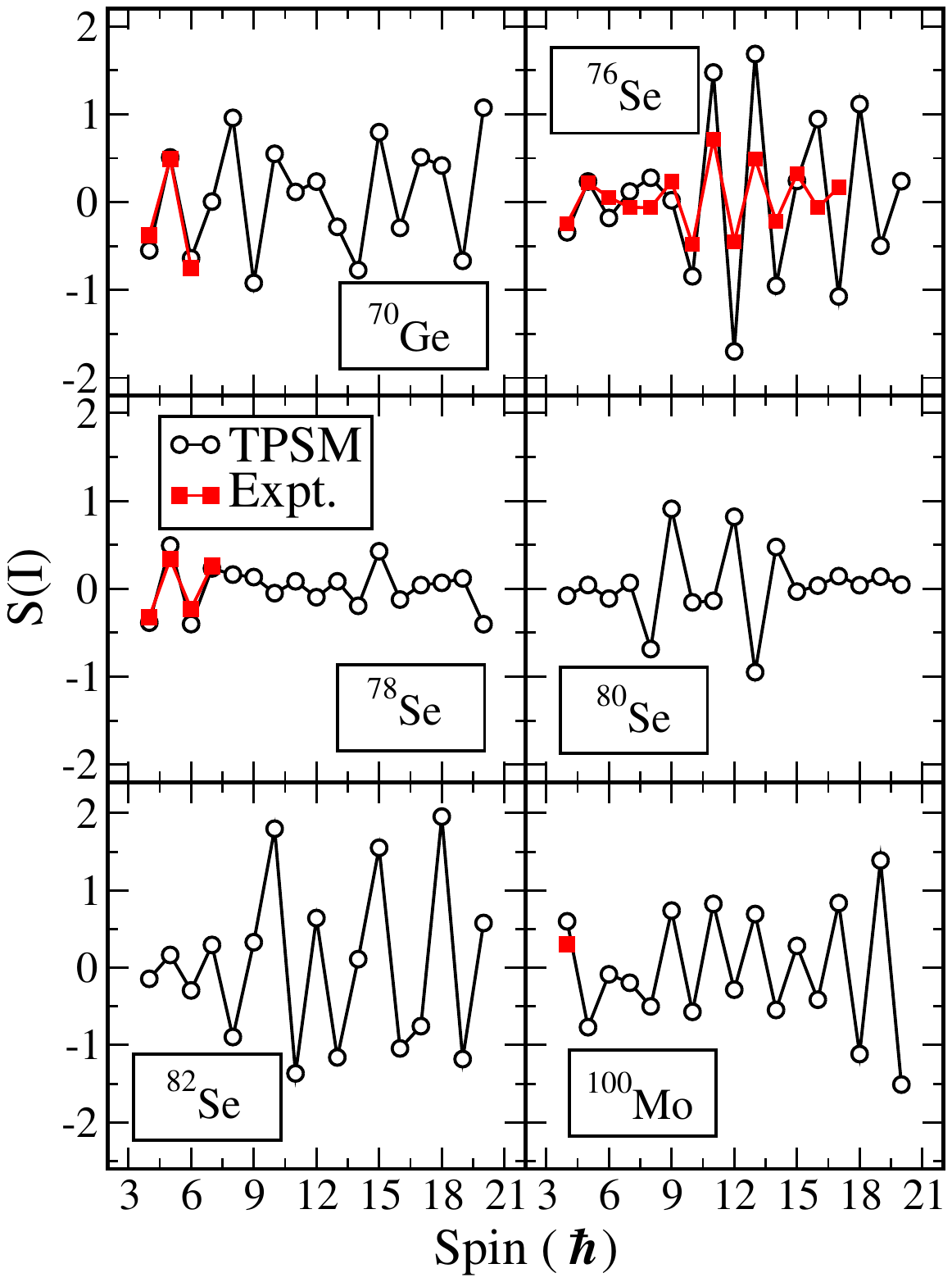}} \caption{(Color
online) Staggering parameters $S(I)$ of the $\gamma$ band in $^{70}$Ge, $^{76,78,80,82}$Se, and $^{100}$Mo.
  }
\label{Stag}
\end{figure}
\begin{figure}[htb]
 \centerline{\includegraphics[trim=0cm 0cm 0cm
0cm,width=0.5\textwidth,clip]{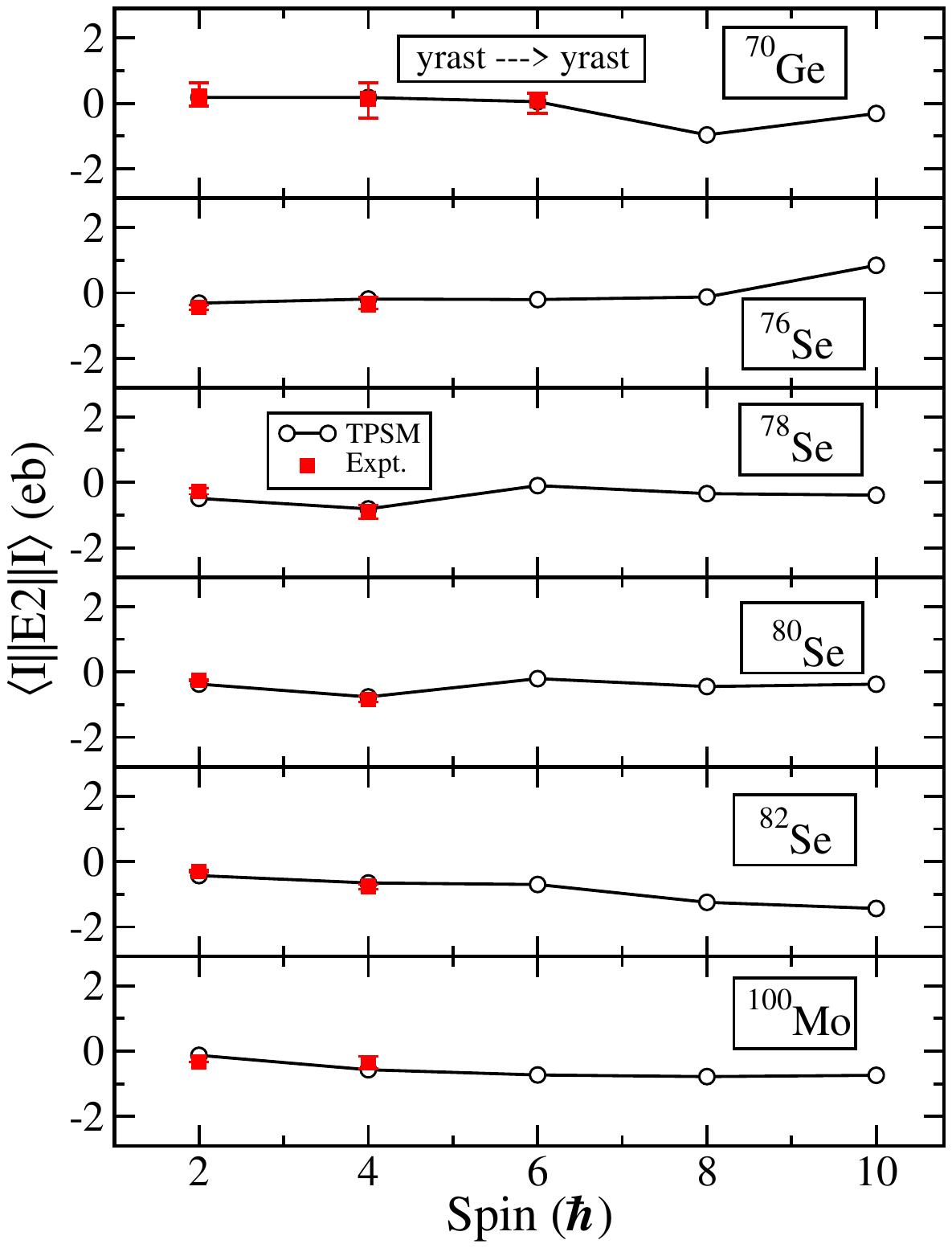}} \caption{(Color
online) Reduced diagonal $E2$ matrix elements for the yrast states in $^{70}$Ge, $^{76,78,80,82}$Se, and $^{100}$Mo isotopes. Expt. data is taken from the Refs. \cite{{cmks_jst1_70ge,KAVKA1995177,Hayakawa2003,Wrzosek064305}}.
  }
\label{E2-dia-y}
\end{figure}
\begin{figure}[htb]
 \centerline{\includegraphics[trim=0cm 0cm 0cm
0cm,width=0.5\textwidth,clip]{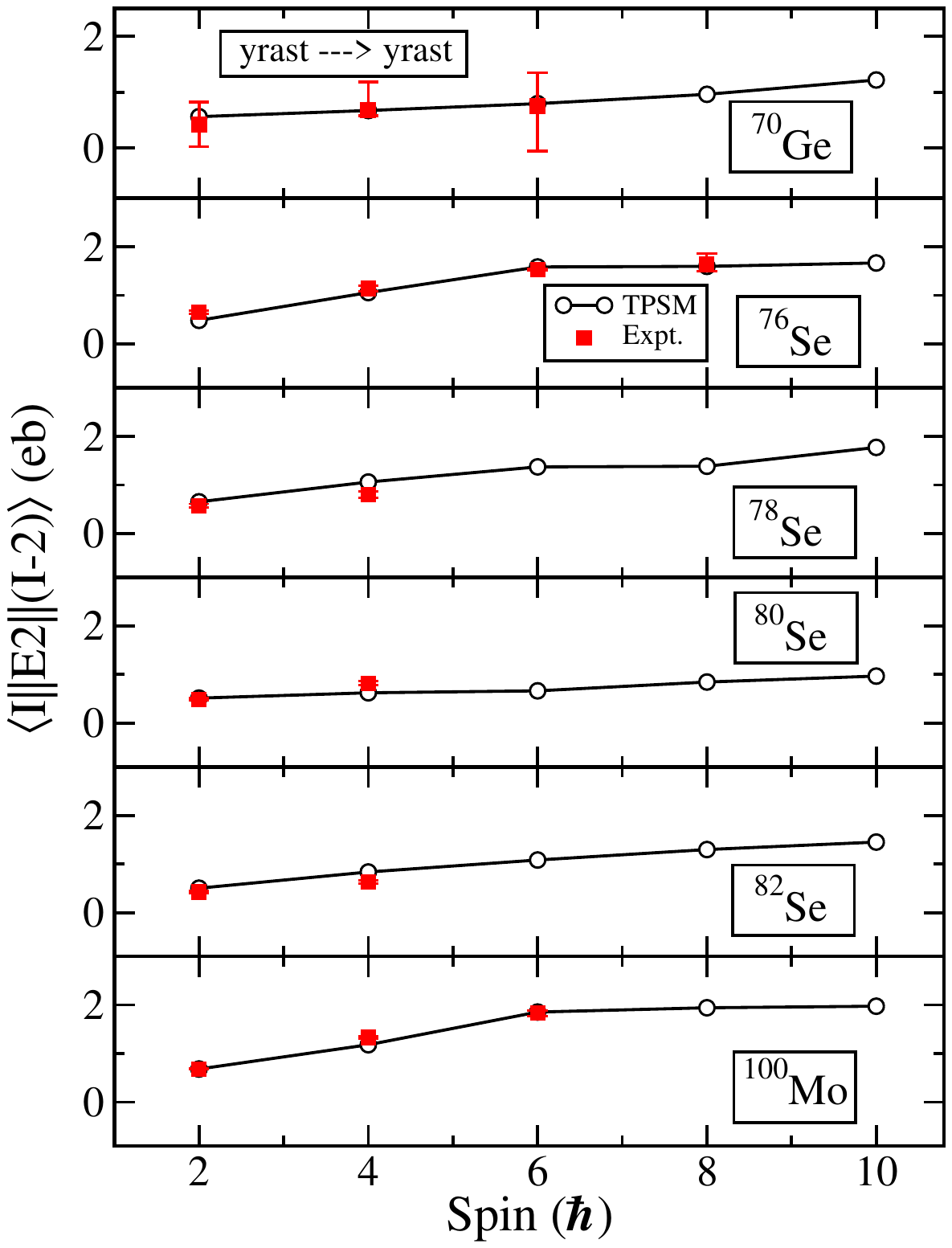}} \caption{(Color
online) Reduced  in-band $E2$ matrix elements for transitions yrast $\rightarrow$ yrast in $^{70}$Ge, $^{76,78,80,82}$Se, and $^{100}$Mo isotopes. Expt. data is taken from the Refs. \cite{{cmks_jst1_70ge,KAVKA1995177,Hayakawa2003,Wrzosek064305}}.
  }
\label{E22-in-y}
\end{figure}
\begin{figure}[t]
 \centerline{\includegraphics[trim=0cm 0cm 0cm
0cm,width=0.5\textwidth,clip]{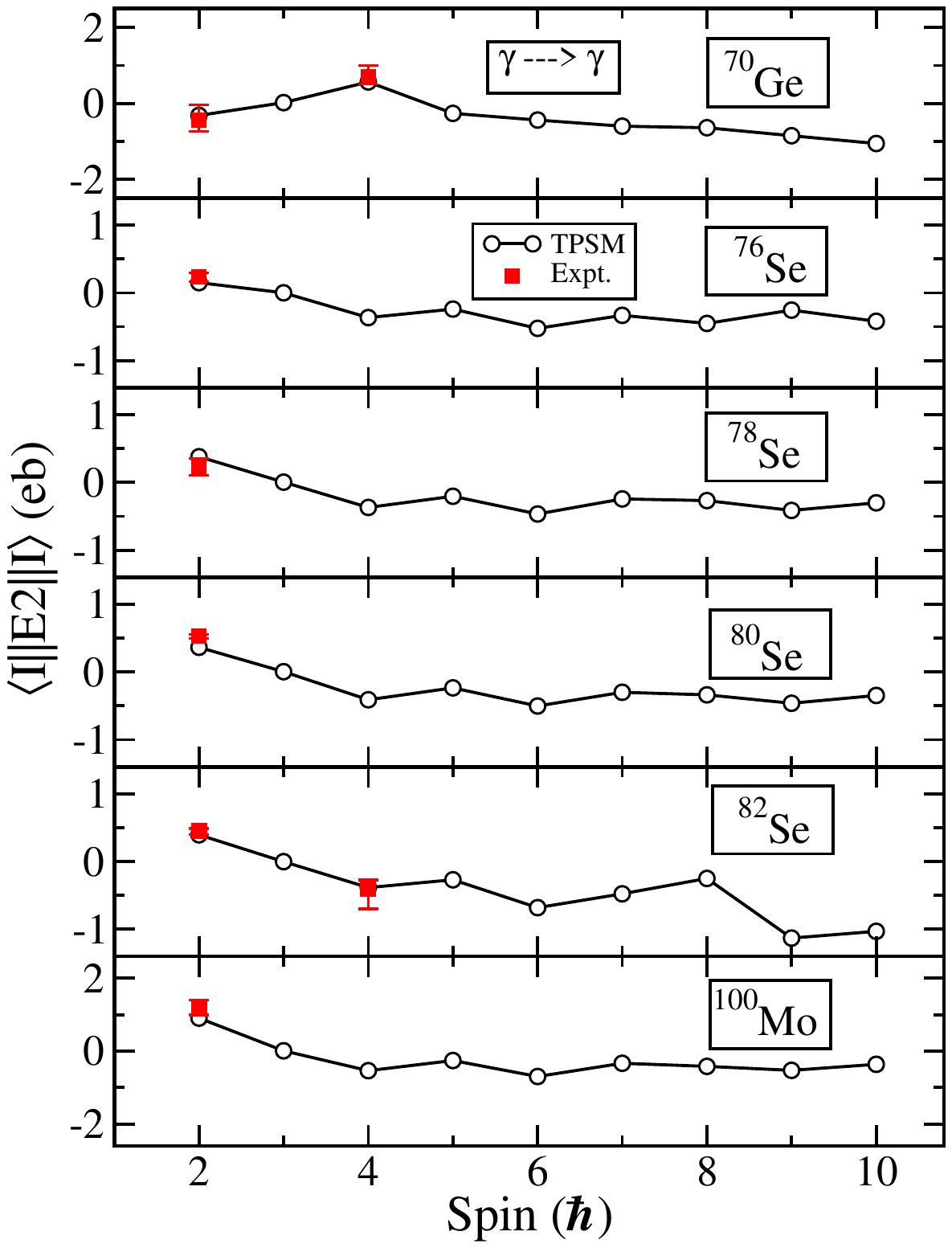}} \caption{(Color
online) Reduced diagonal $E2$ matrix elements  for the $\gamma$ band states $^{70}$Ge, $^{76,78,80,82}$Se, and $^{100}$Mo isotopes. Expt. data is taken from the Refs. \cite{{cmks_jst1_70ge,KAVKA1995177,Hayakawa2003,Wrzosek064305}}.
  }
\label{E2-dia-g}
\end{figure}
\begin{figure}[t]
 \centerline{\includegraphics[trim=0cm 0cm 0cm
0cm,width=0.5\textwidth,clip]{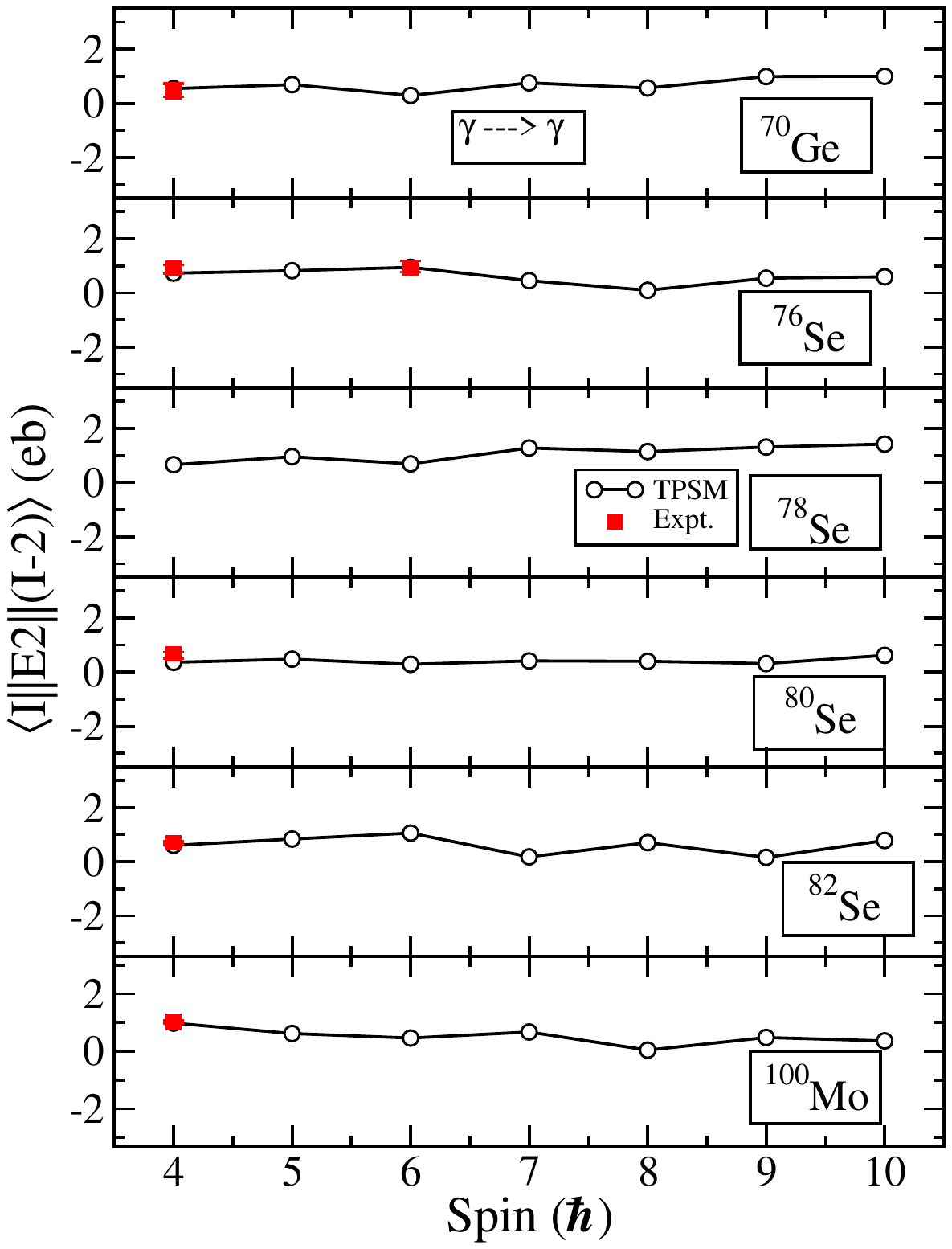}} \caption{(Color
online) Reduced in-band $E2$ matrix elements for transitions $\gamma$ $\rightarrow$ $\gamma$ in $^{70}$Ge, $^{76,78,80,82}$Se, and $^{100}$Mo isotopes. Expt. data is taken from the Refs. \cite{{cmks_jst1_70ge,KAVKA1995177,Hayakawa2003,Wrzosek064305}}.
  }
\label{E22-in-g}
\end{figure}
\begin{figure}[t]
 \centerline{\includegraphics[trim=0cm 0cm 0cm
0cm,width=0.5\textwidth,clip]{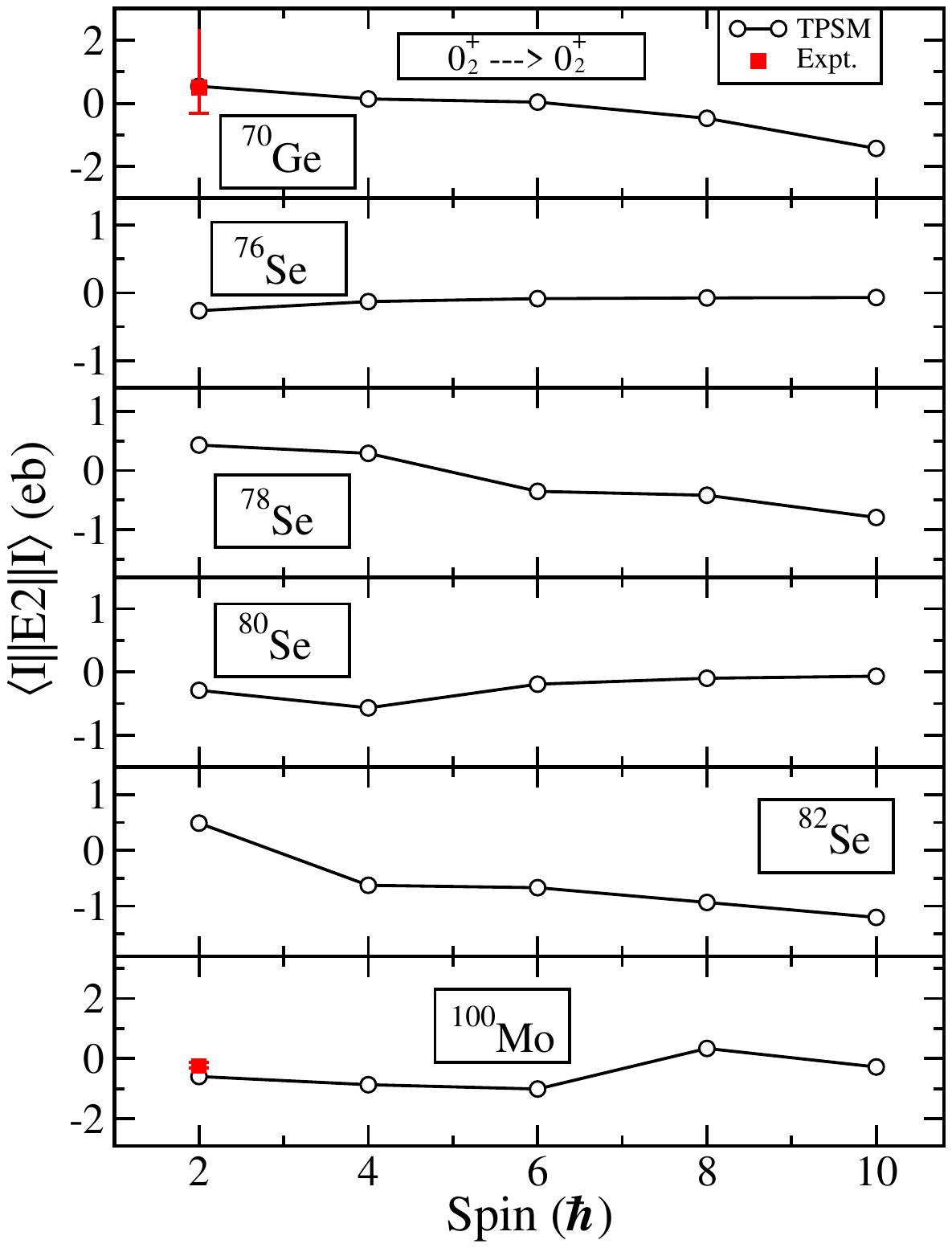}} \caption{(Color
online) Reduced diagonal $E2$ matrix elements  for the $0_2^+$-band states $^{70}$Ge, $^{76,78,80,82}$Se, and $^{100}$Mo isotopes. Expt. data is taken from the Refs. \cite{{cmks_jst1_70ge,KAVKA1995177,Hayakawa2003,Wrzosek064305}}.
  }
\label{E2-dia-0ex}
\end{figure}
\begin{figure}[htb]
 \centerline{\includegraphics[trim=0cm 0cm 0cm
0cm,width=0.5\textwidth,clip]{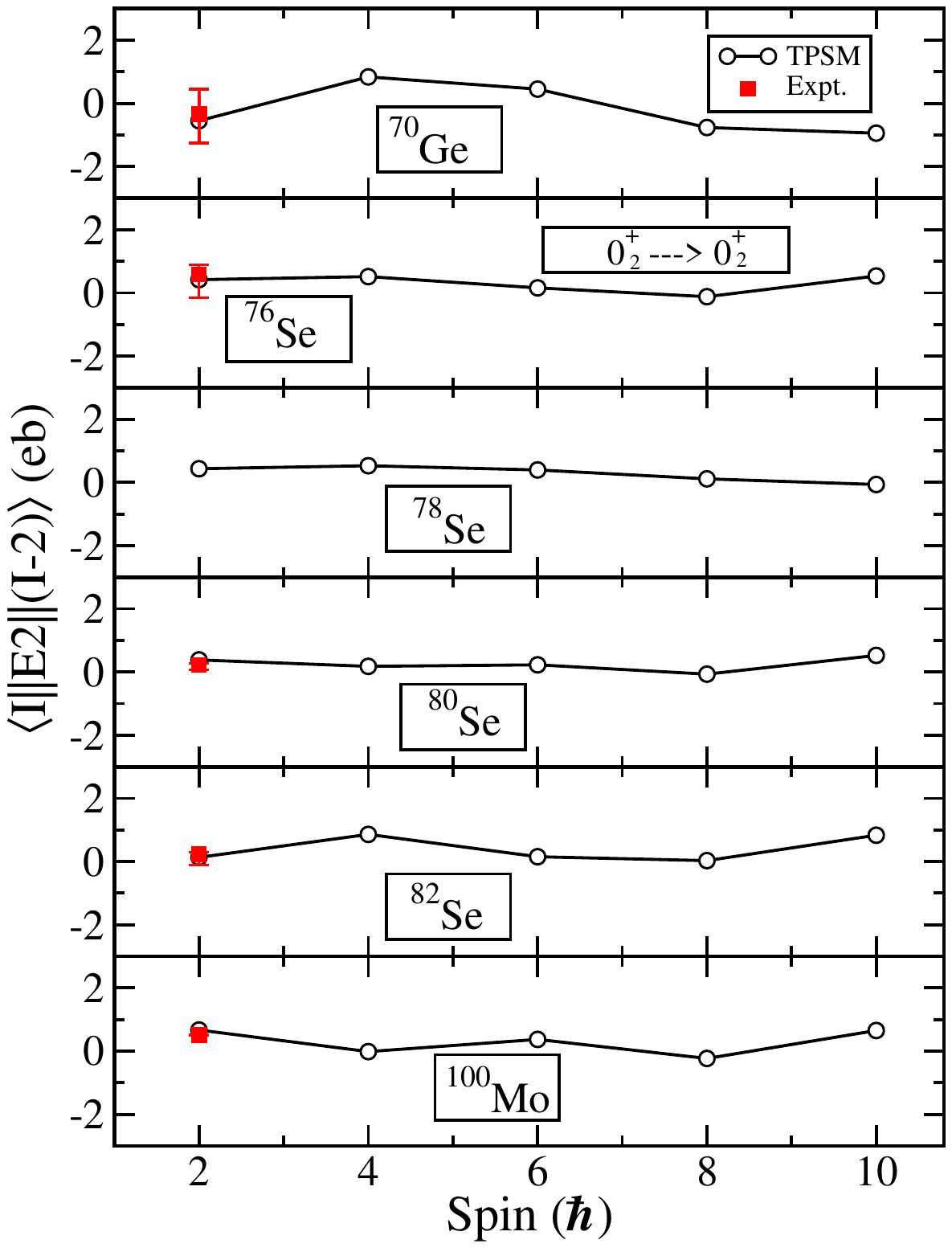}} \caption{(Color
online) Reduced in-band $E2$ matrix elements for transitions $0^+_2$ $\rightarrow$ $0^+_2$ in $^{70}$Ge, $^{76,78,80,82}$Se, and $^{100}$Mo isotopes. Expt. data is taken from the Refs. \cite{{cmks_jst1_70ge,KAVKA1995177,Hayakawa2003,Wrzosek064305}}.
  }
\label{E22-ex0}
\end{figure}
\begin{figure}[htb]
 \centerline{\includegraphics[trim=0cm 0cm 0cm
0cm,width=0.5\textwidth,clip]{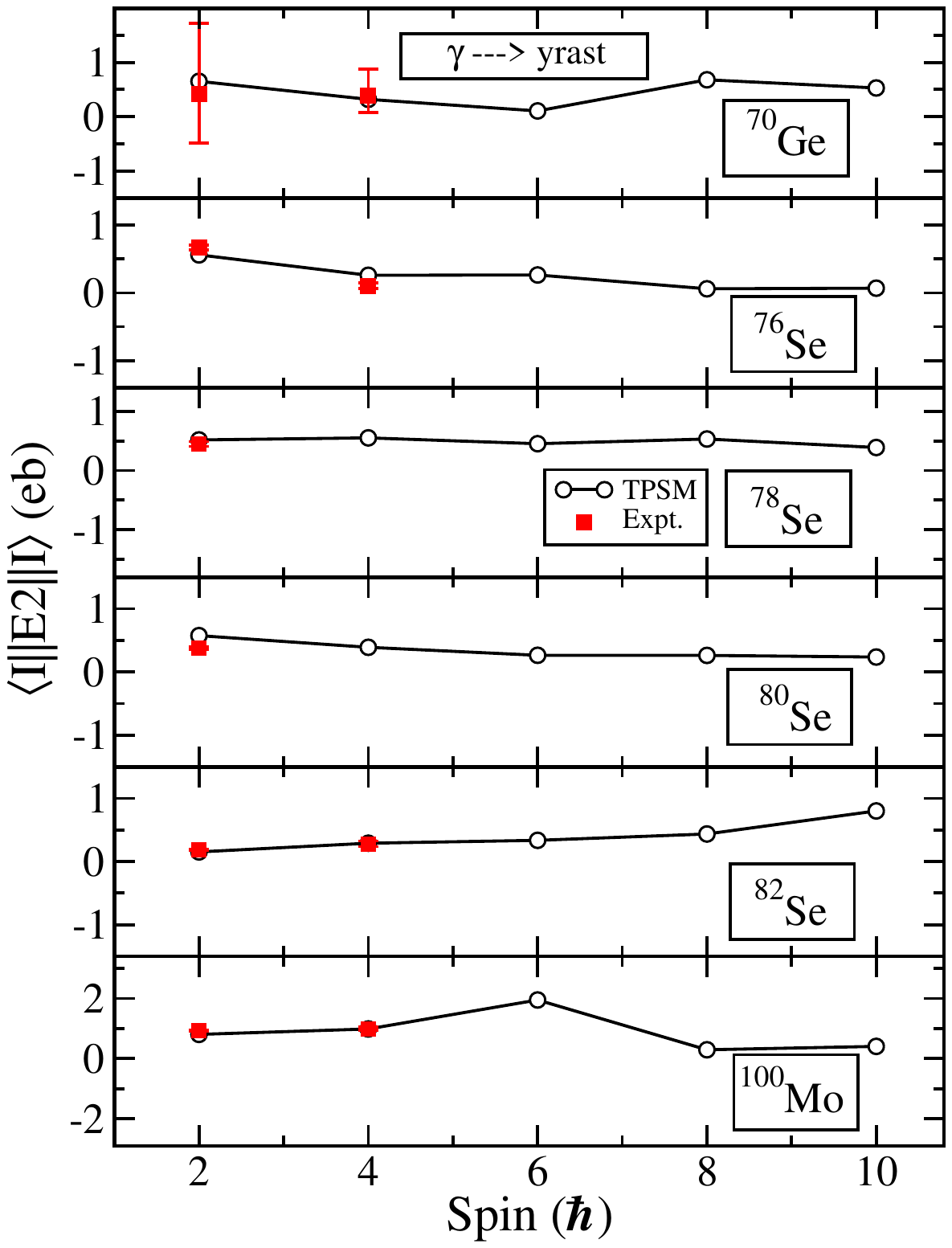}} \caption{(Color
online) Reduced inter-band $E2$ matrix elements for transitions $\gamma$ $\rightarrow$ yrast in $^{70}$Ge, $^{76,78,80,82}$Se and $^{100}$Mo isotopes. Expt. data is taken from the Refs. \cite{{cmks_jst1_70ge,KAVKA1995177,Hayakawa2003,Wrzosek064305}}.
  }
\label{E20-gy}
\end{figure}
\begin{figure}[htb]
 \centerline{\includegraphics[trim=0cm 0cm 0cm
0cm,width=0.5\textwidth,clip]{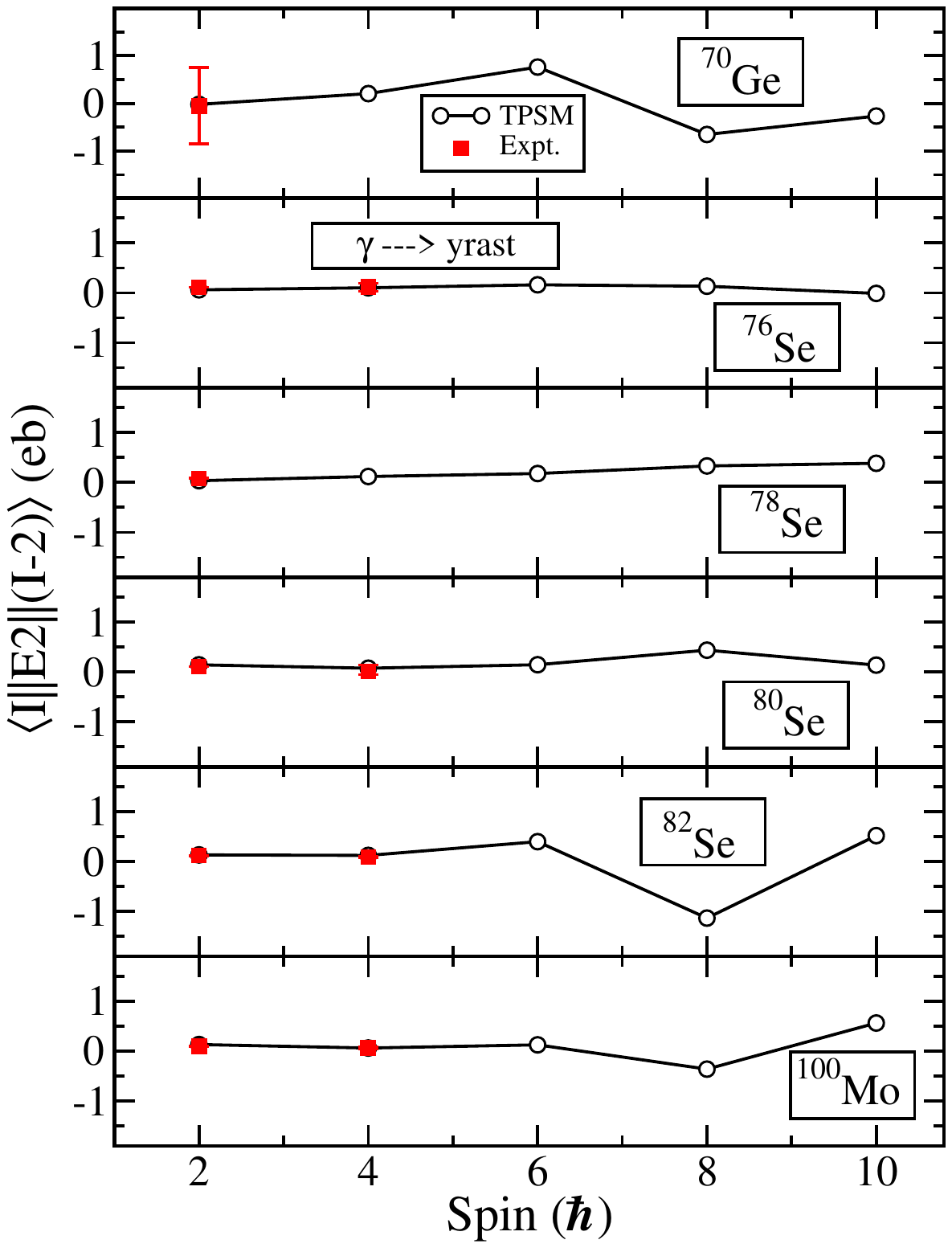}} \caption{(Color
online) Reduced inter-band $E2$ matrix elements for transitions $\gamma$ $\rightarrow$ yrast in $^{70}$Ge, $^{76,78,80,82}$Se, and $^{100}$Mo isotopes. Expt. data is taken from the Refs. \cite{{cmks_jst1_70ge,KAVKA1995177,Hayakawa2003,Wrzosek064305}}.
  }
\label{E22-gy}
\end{figure}
\begin{figure}[!h]
  \centerline{\includegraphics[trim=0cm 0cm 0cm
0cm,width=0.5\textwidth,clip]{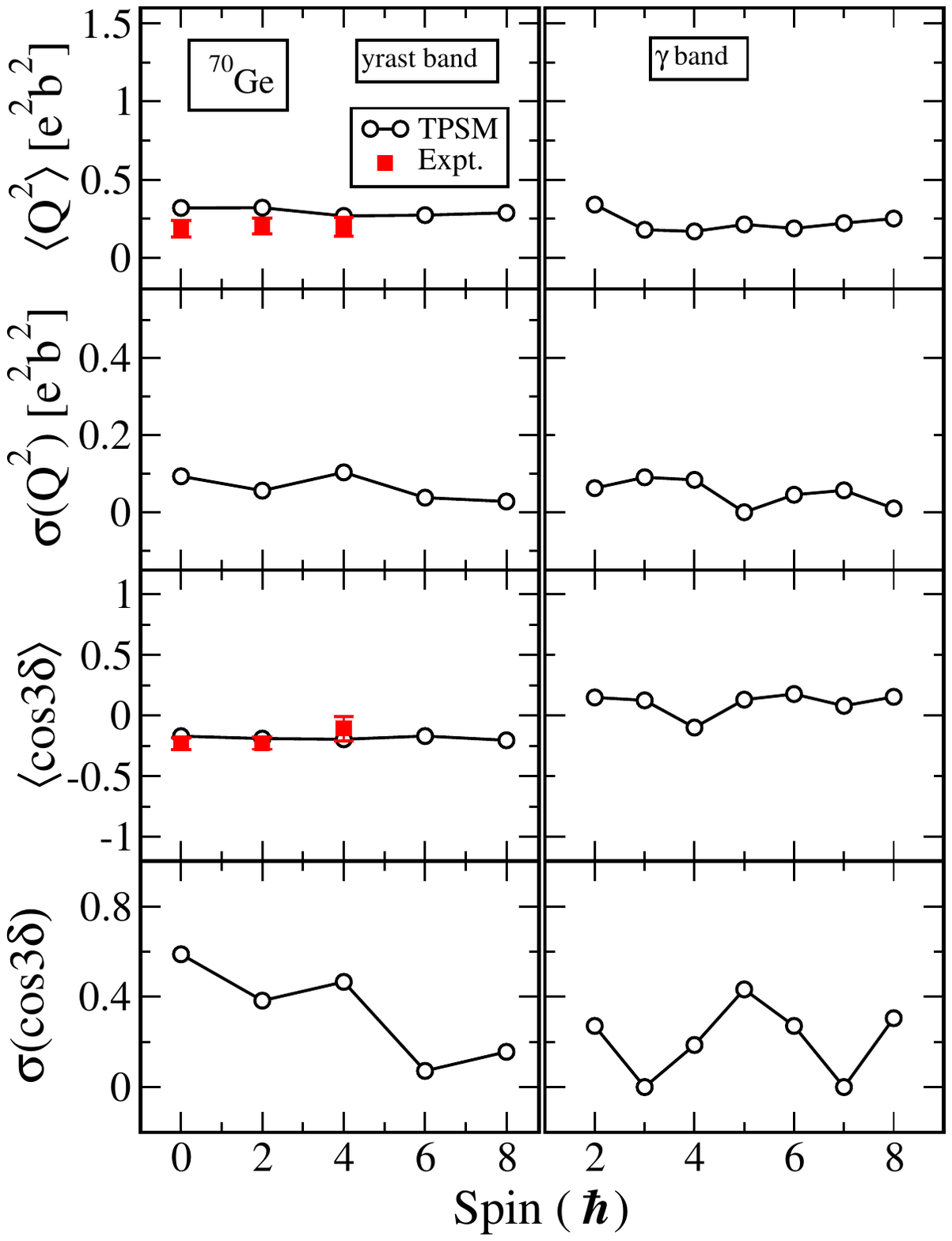}} \caption{(Color
online) Centroid $\langle Q^2 \rangle$, dispersion $ \sigma(Q^2)$, centroid $\langle \textrm{cos}3\delta \rangle$ and
dispersion $\sigma(\textrm {cos}3\delta)$ for the yrast and $\gamma$ bands in $^{70}$Ge. }
\label{f1:sigmaQ2}
\end{figure}
\begin{figure}[!h]
  \centerline{\includegraphics[trim=0cm 0cm 0cm
0cm,width=0.5\textwidth,clip]{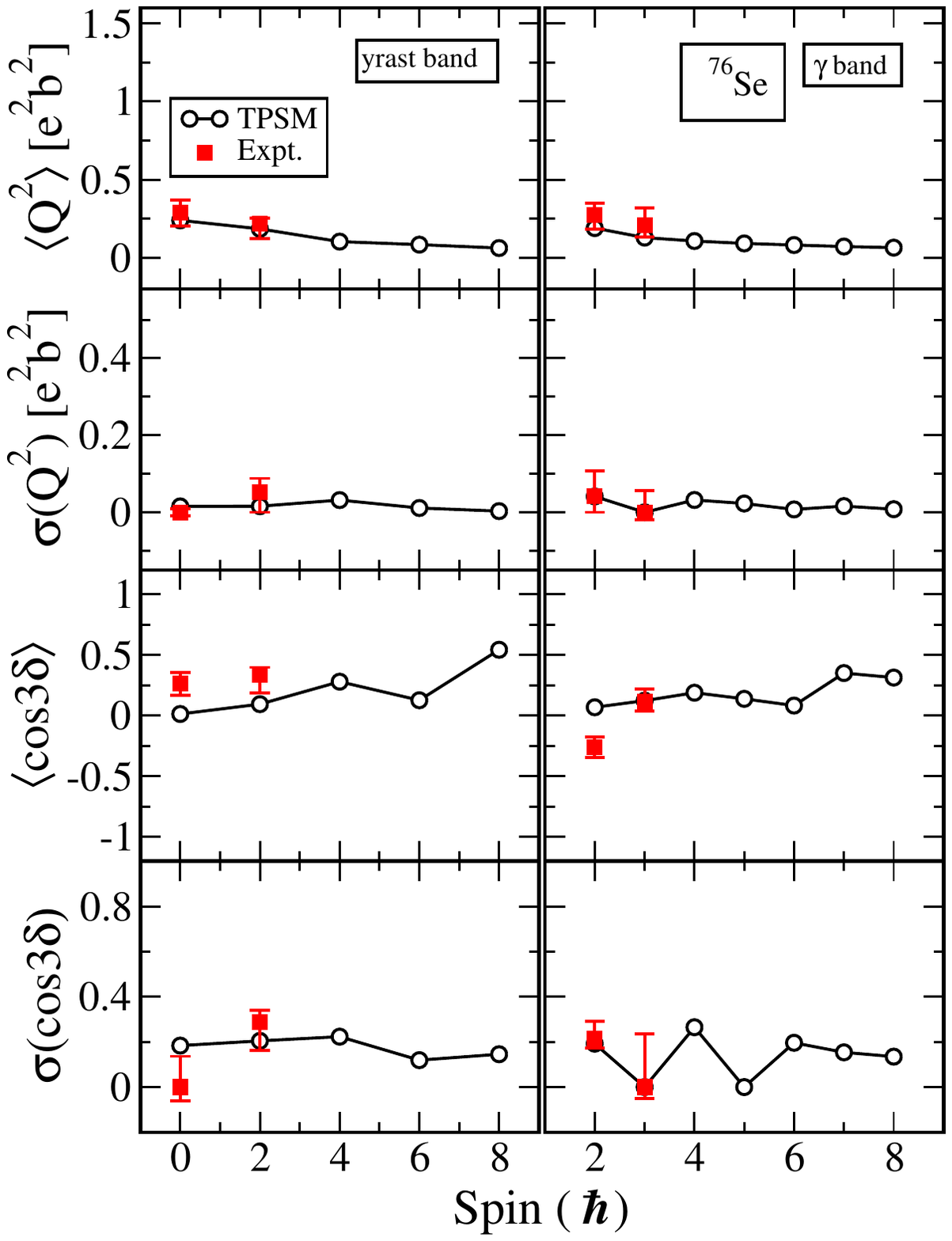}} \caption{(Color
online) Centroid $\langle Q^2 \rangle$, dispersion $ \sigma(Q^2)$, centroid $\langle \textrm{cos}3\delta \rangle$ and
dispersion $\sigma(\textrm {cos}3\delta)$ for the yrast and $\gamma$ bands in $^{76}$Se. }
\label{f2:sigmaQ2}
\end{figure}
\begin{figure}[!h]
  \centerline{\includegraphics[trim=0cm 0cm 0cm
0cm,width=0.5\textwidth,clip]{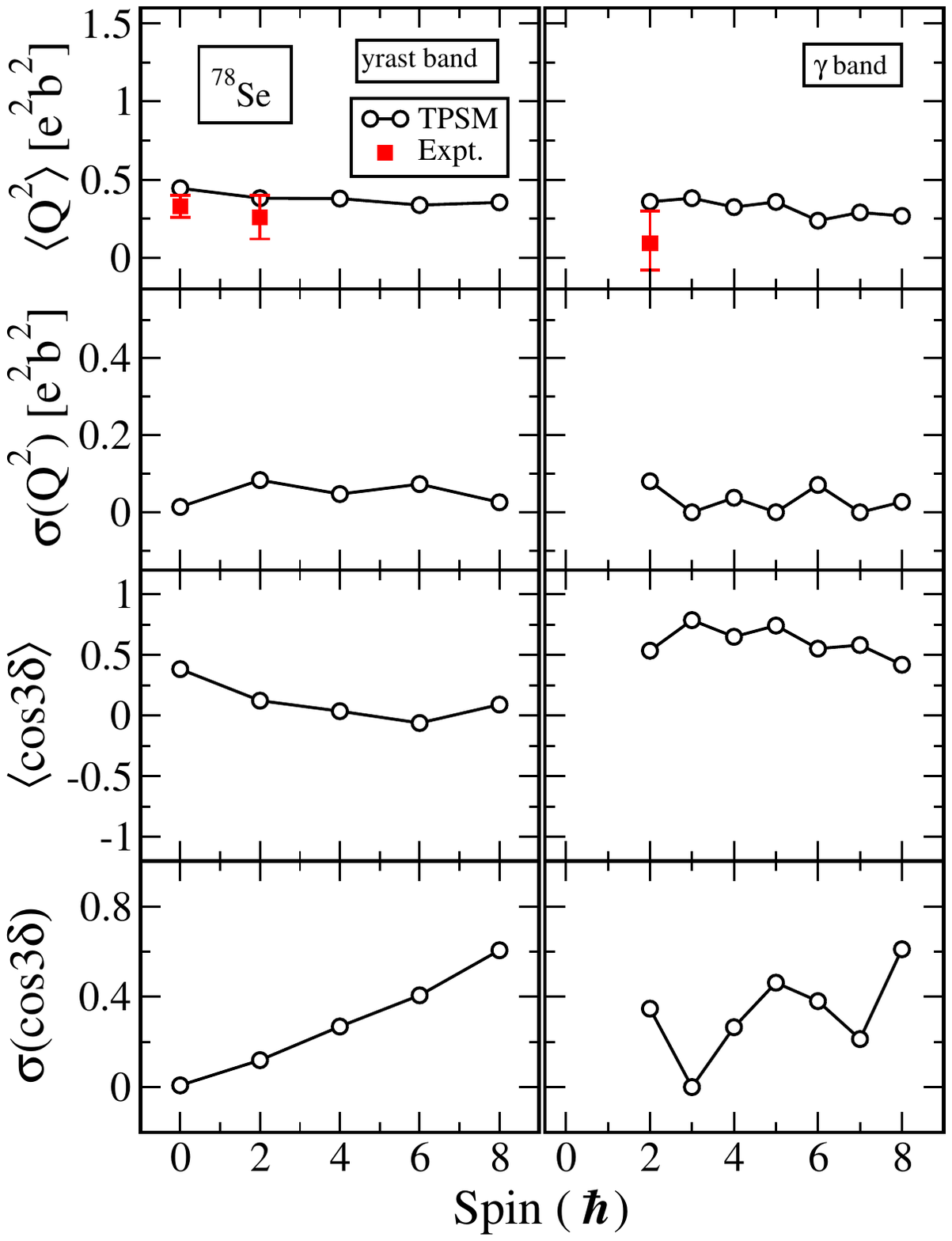}} \caption{(Color
online) Centroid $\langle Q^2 \rangle$, dispersion $ \sigma(Q^2)$, centroid $\langle \textrm{cos}3\delta \rangle$ and
dispersion $\sigma(\textrm {cos}3\delta)$ for the yrast and $\gamma$ bands in $^{78}$Se. }
\label{f3:sigmaQ2}
\end{figure}
\begin{figure}[!h]
  \centerline{\includegraphics[trim=0cm 0cm 0cm
0cm,width=0.5\textwidth,clip]{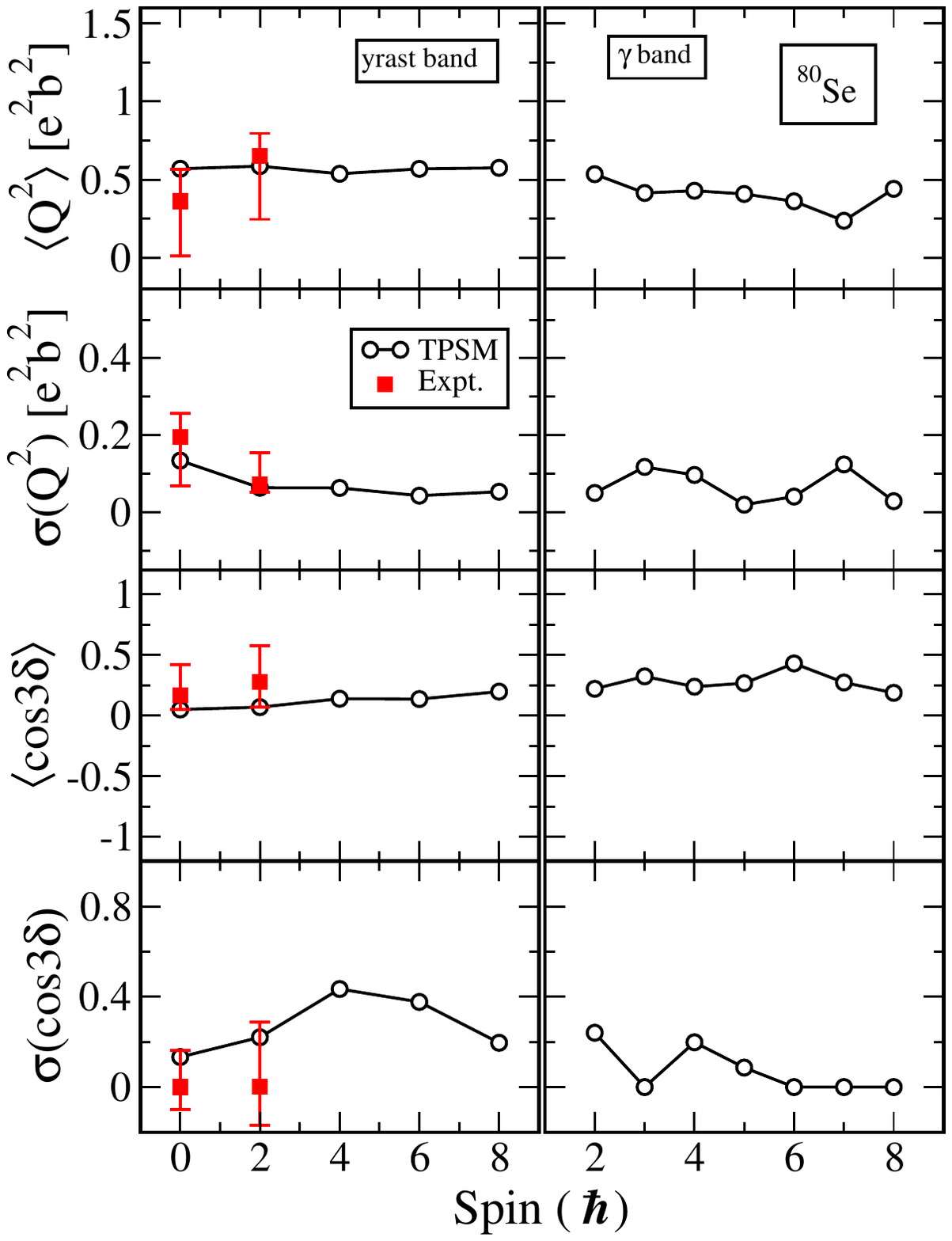}} \caption{(Color
online) Centroid $\langle Q^2 \rangle$, dispersion $ \sigma(Q^2)$, centroid $\langle \textrm{cos}3\delta \rangle$ and
dispersion $\sigma(\textrm {cos}3\delta)$ for the yrast and $\gamma$ bands in$^{80}$Se. }
\label{f4:sigmaQ2}
\end{figure}
\begin{figure}[!h]
  \centerline{\includegraphics[trim=0cm 0cm 0cm
0cm,width=0.5\textwidth,clip]{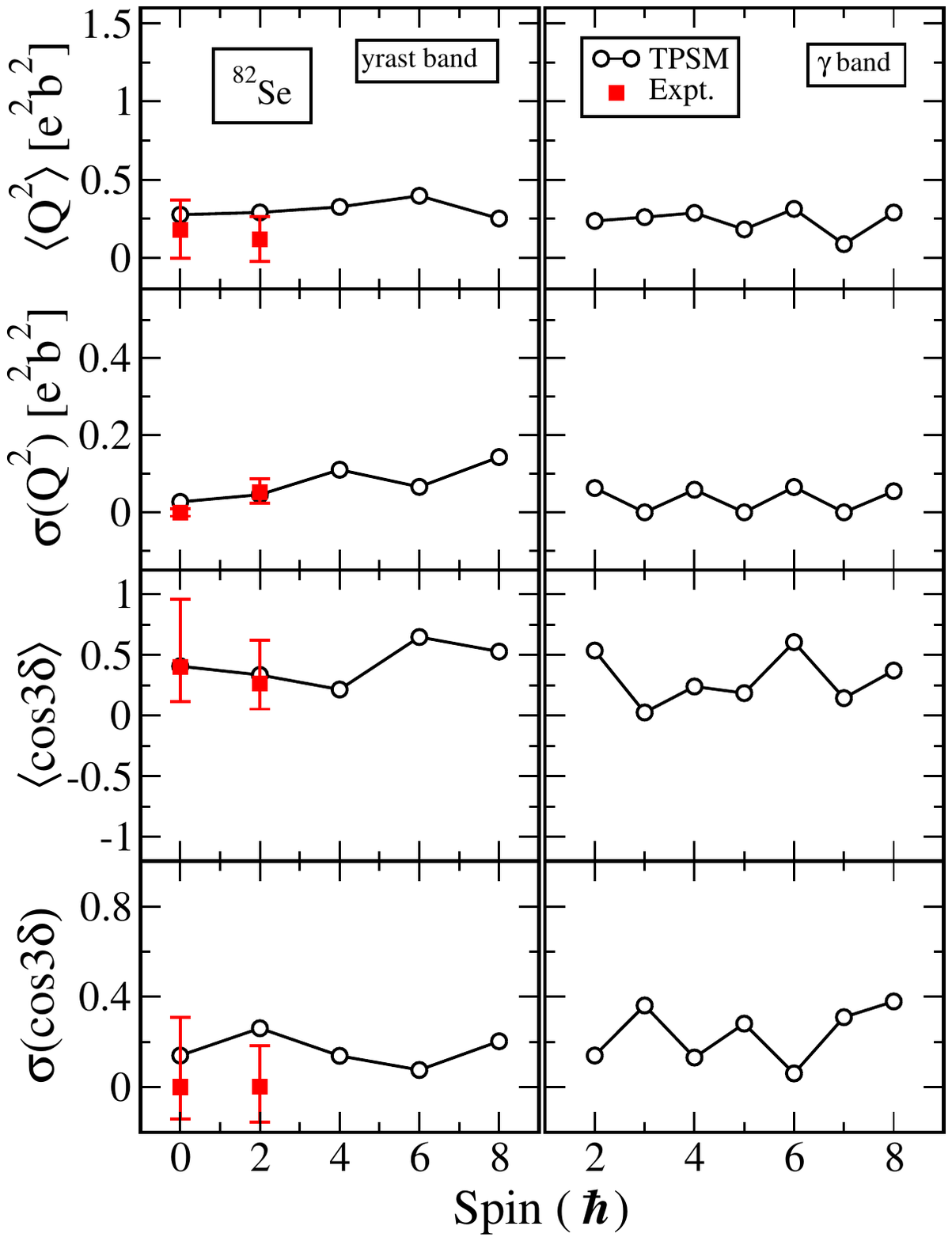}} \caption{(Color
online) Centroid $\langle Q^2 \rangle$, dispersion $ \sigma(Q^2)$, centroid $\langle \textrm{cos}3\delta \rangle$ and
dispersion $\sigma(\textrm {cos}3\delta)$ for the yrast and $\gamma$ bands in $^{82}$Se. }
\label{f5:sigmaQ2}
\end{figure}
\begin{figure}[!h]
  \centerline{\includegraphics[trim=0cm 0cm 0cm
0cm,width=0.5\textwidth,clip]{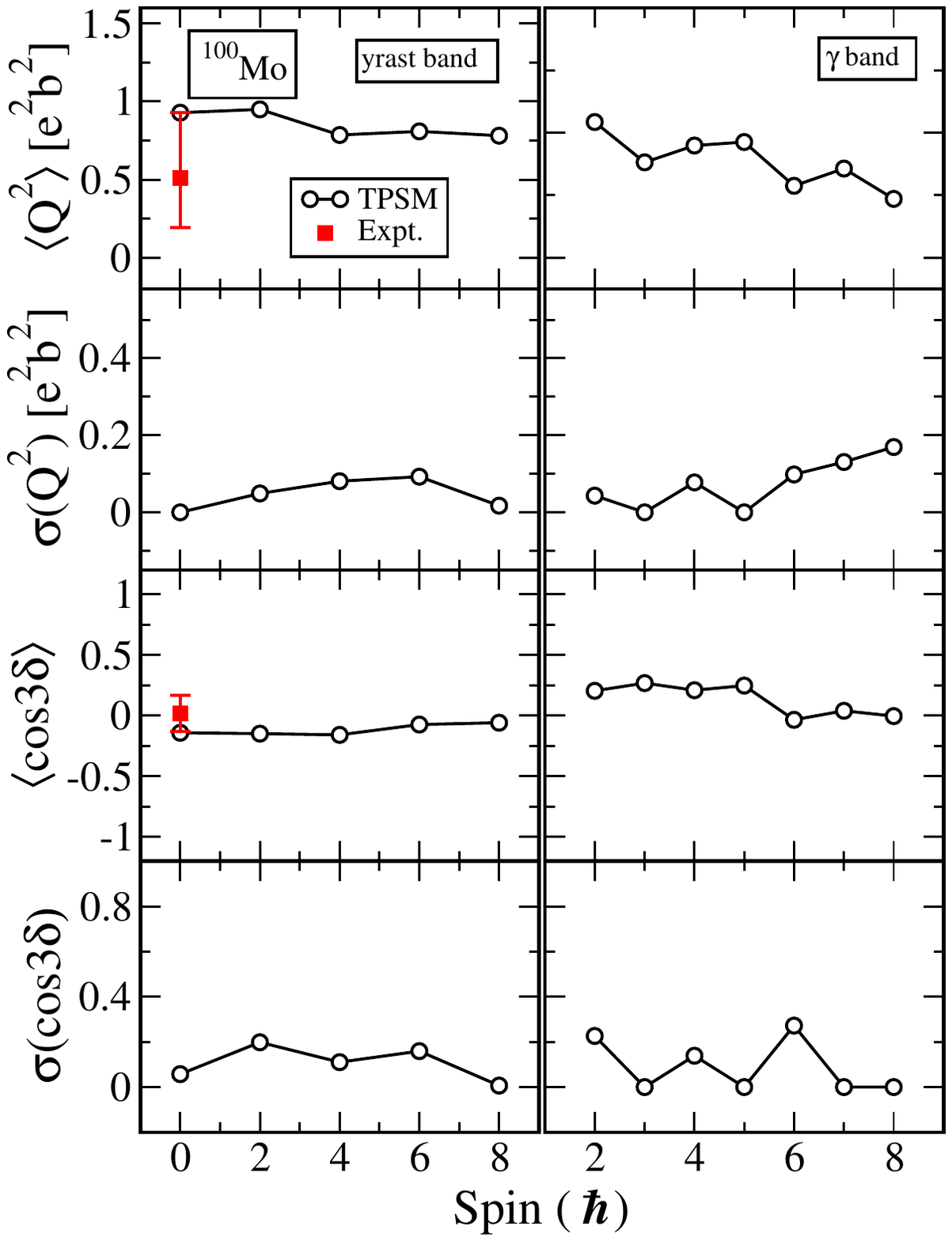}} \caption{(Color
online)Centroid $\langle Q^2 \rangle$, dispersion $ \sigma(Q^2)$, centroid $\langle \textrm{cos}3\delta \rangle$ and
dispersion $\sigma(\textrm {cos}3\delta)$ for the yrast and $\gamma$ bands in $^{100}$Mo. }
\label{f6:sigmaQ2}
\end{figure}
 \begin{figure}[!h]
 \centerline{\includegraphics[trim=0cm 0cm 0cm
 0cm,width=0.5\textwidth,clip]{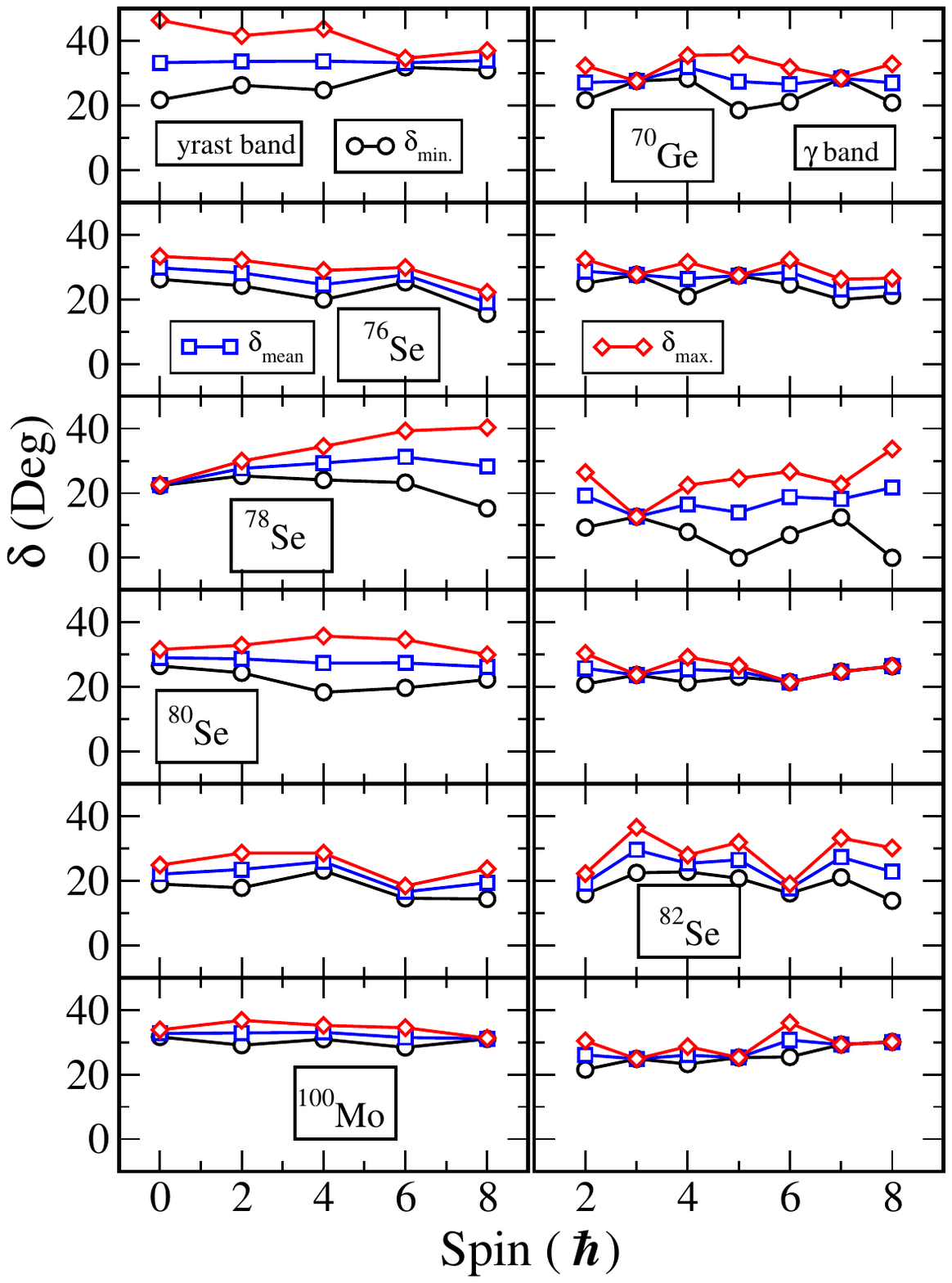}} \caption{(Color
 online) Bands of the triaxiality parameter $\delta$ that contain 68\% of the probability, assuming  normal distributions,
 with the dispersions $\sigma(\textrm {cos}3\delta)$ and the centroids $\langle \textrm{cos}3\delta \rangle$ shown 
 in Figs. \ref{f1:sigmaQ2}-\ref{f6:sigmaQ2}.}
 \label{f7:delta_bands}
\end{figure}

In recent works \cite{nazira,rouoofe2}, we performed a detailed analysis of the COULEX data of nine nuclides of
$^{72}$Ge, $^{76}$Ge, $^{104}$Ru, $^{168}$Er, $^{186}$Os, $^{188}$Os, $^{190}$Os, $^{192}$Os, and $^{194}$Pt. The analysis was carried out using the
triaxial projected shell model (TPSM) approach, and it was demonstrated that this model provided a good description of the large sets
of $E2$ matrix elements known for the nine nuclides from the COULEX experiments \cite{Kotlinski1990,Ayangeakaa2016,DC86,Wu1996,104Ru,Ayangeakaa2019}. 
It was also shown that quasiparticle mixing into the $\gamma$ rigid vacuum configuration
was responsible for reproducing  $\gamma$ soft features as deduced from the measured $E2$ matrix elements of
the studied nuclides \cite{JS21,Rouoof2024}. The purpose of the present investigation
is to extend the TPSM analysis of $E2$ matrix elements for nuclides that have become recently available from COULEX experiments, and also to those nuclides that
were inadvertently omitted in our previous study \cite{rouoofe2}.  Clearly, COULEX experiments have been performed for a wide range of nuclei,
but we have chosen only those systems for which large set of $E2$ matrix elements are available such that shape invariant analysis can be performed.

In the present work, we have studied $^{70}$Ge, $^{76,78,80,82}$Se and $^{100}$Mo nuclides. $^{70}$Ge has been recently studied
through the multi-step Coulomb excitation experiment and twenty-seven transitional and six diagonal matrix have been deduced from the
measured cross sections \cite{cmks_jst1_70ge}. For $^{76,80,82}$Se isotopes, COULEX experiment with $^{16}$O, $^{48}$Ti and
$^{208}$Pb beams determined about twenty $E2$ matrix elements for each of the three systems \cite{KAVKA1995177}. In the projectile
Coulomb excitation experiment \cite{Hayakawa2003}, nine $E2$ matrix elements of $^{78}$Se were determined. $^{100}$Mo was studied with 76 MeV
$^{32}$Si beam and twenty-six  $E1$, $E2$, $E3$, and $M1$ matrix elements were determined \cite{Wrzosek064305}.
The TPSM approach has been used to perform the detailed analysis
of excitation energies, $E2$ matrix elements and shape invariant quantities of these six nuclei. The present manuscript is organized
in the following manner. In the next section \ref{TPSM}, the TPSM approach is briefly
described for completeness. In section \ref{Results}, the results of excitation energies, $E2$ matrix elements and rotational invariant quantities
are presented and discussed. Finally, the present work is summarized and concluded in section \ref{Sum_Con}.

\section{ TRIAXIAL PROJECTED SHELL MODEL APPROACH } \label{TPSM}

TPSM is a shell model approach with a configuration  space composed of angular momentum projected deformed basis rather than the spherical one's \cite{SH99}.
This model employs  the pairing
plus quadrupole Hamiltonian (PPQM) of Kumar and Baranger \cite{Kumar1968} and performs three dimensional angular momentum
projection of deformed triaxial configurations. TPSM is a microscopic framework with the collective behavior
emerging from the solution of Hamiltonian expressed in terms of the nucleonic degrees of freedom.
In recent years, the TPSM approach has emerged as tool of choice to provide a unified
description of the spectroscopic properties of deformed and transitional
nuclei \cite{SH99,JS16,JS21,Jeh2022,nazira22,nazira,NaziraNP,Rouoof2024,rouoofe2} up to quite high spin states.

The starting point in the TPSM calculation is the construction of the quasiparticle deformed basis states, which are obtained by
solving the triaxially deformed Nilsson Hamiltonian and then incorporating the pairing correlations within the BCS approximation. The
triaxial Nilsson Hamiltonian used in the calculations is given by
\begin{equation}
\hat{H}_{N}
=
\hat{H}_{0}
-
\frac{2}{3}\hbar\omega
\left[
\varepsilon \hat{Q}_{0}
+
\varepsilon^{\prime}
\frac{\hat{Q}_{+2}+\hat{Q}_{-2}}{\sqrt{2}}
\right],
\end{equation}
where $\hat{H}_{0}$ denotes the spherical single-particle Hamiltonian that includes the spin-orbit interaction. The parameters
$\varepsilon$ and $\varepsilon^{\prime}$ represent the axial and non-axial quadrupole deformations, respectively, and correspond to the triaxiality parameter
\begin{equation}
\gamma = \tan^{-1}(\varepsilon^{\prime}/\varepsilon).
\end{equation}
These deformation parameters are treated as fixed inputs and are chosen to reproduce the experimental observables such as the excitation energy of the $\gamma$ band head and the reduced transition probability $B(E2;2_1^+ \rightarrow 0_1^+)$~\cite{JS21}.

Since the intrinsic quasiparticle states violate the rotational symmetry, the restoration of good angular momentum basis is achieved through three-dimensional angular-momentum projection operators, given by
\begin{equation}
\hat{P}^{I}_{MK}
=
\frac{2I+1}{8\pi^{2}}
\int d\Omega\,
D^{I}_{MK}(\Omega)\,
\hat{R}(\Omega),
\end{equation}
where $D^{I}_{MK}(\Omega)$ are Wigner $D$ functions and $\hat{R}(\Omega)$ denotes the rotation operator defined in terms of Euler angles.

For even-even nuclei, the TPSM model space consists of angular-momentum projected configurations built from the triaxial vacuum and from quasiparticle excitations. The basis space is given by
\begin{align}
\{
\hat{P}^{I}_{MK}|\Phi\rangle,\;
&\hat{P}^{I}_{MK}a^{\dagger}_{p_1}a^{\dagger}_{p_2}|\Phi\rangle,\nonumber\\
&\hat{P}^{I}_{MK}a^{\dagger}_{n_1}a^{\dagger}_{n_2}|\Phi\rangle,\;
\hat{P}^{I}_{MK}a^{\dagger}_{p_1}a^{\dagger}_{p_2}
a^{\dagger}_{n_1}a^{\dagger}_{n_2}|\Phi\rangle
\}.
\end{align}

The above projected basis states are then employed to diagonalize the pairing plus quadrupole--quadrupole Hamiltonian :
\begin{equation}
\hat{H}
=
\hat{H}_{0}
-
\frac{1}{2}\chi\sum_{\mu}\hat{Q}^{\dagger}_{\mu}\hat{Q}_{\mu}
-
G_{M}\hat{P}^{\dagger}\hat{P}
-
G_{Q}\sum_{\mu}\hat{P}^{\dagger}_{\mu}\hat{P}_{\mu},
\end{equation}
where $\chi$ denotes the quadrupole--quadrupole interaction strength and $G_{M}$ and $G_{Q}$ represent the monopole and quadrupole pairing strengths. These interaction strengths are chosen consistently with established TPSM systematics and constrained by the empirical pairing gaps and deformation properties~\cite{JS21,SH99}.

As the projected states are not orthonormal, Hill-Wheeler approach is employed which gives rise to a generalized eigenvalue equation. The resulting
TPSM eigenstate with good angular momentum is expressed as
\begin{equation}
|\Psi^{\sigma}_{IM}\rangle
=
\sum_{\kappa,K}
f^{\sigma I}_{\kappa K}
\hat{P}^{I}_{MK}
|\Phi_{\kappa}\rangle ,
\end{equation}
where $f^{\sigma I}_{\kappa K}$ denote the configuration-mixing amplitudes. 

\begin{table*}[htp!]
\caption{\label{tab:parameters} Axial and triaxial quadrupole deformation parameters,
  $\epsilon$ and $\epsilon'$,  employed in the TPSM calculation. Axial deformations are taken from the earlier
  works \cite{JS21,Rouoof2024} and \cite{raman},
and nonaxial deformations are chosen in such a way that heads
of the $\gamma$ bands are reproduced.}
\begin{tabular}
{p{2.4cm}p{2.4cm}p{2.4cm}p{2.4cm}p{2.4cm}p{2.4cm}p{2.4cm}p{0.9cm}}\\

  \hline\hline
 Isotope  	&$^{70}$Ge	&$^{76}$Se	&$^{78}$Se	&$^{80}$Se	&$^{82}$Se	&$^{100}$Mo 		\\
  \hline									
$\epsilon$	&0.230	&0.260	&0.256	&0.265	&0.283	&0.230		\\
$\epsilon'$	&0.168	&0.155	&0.150	&0.180	&0.160	&0.190		\\

\hline\hline								

\end{tabular}
\end{table*}

The reduced electric quadrupole transition probability between an initial state
$|\Psi^{\sigma_i}_{I_i}\rangle$ and a final state
$|\Psi^{\sigma_f}_{I_f}\rangle$
is given by
\begin{equation}
B(E2; I_i \rightarrow I_f)
=
\frac{e^{2}}{2I_i+1}
\left|
\langle \Psi^{\sigma_f}_{I_f}
\| \hat{Q}_{2} \|
\Psi^{\sigma_i}_{I_i} \rangle
\right|^{2}.
\end{equation}
Within the TPSM framework, the reduced matrix element contains contributions from both the intrinsic quasiparticle structure and the collective rotational motion generated through angular-momentum projection.
The reduced matrix element of a general irreducible tensor operator $\hat{O}_{L}$ between TPSM states can be written as
\begin{align}
\langle \Psi^{\sigma_f}_{I_f} \| \hat{O}_{L} \| \Psi^{\sigma_i}_{I_i} \rangle
&=
\sum_{\kappa_f,\kappa_i}
\sum_{K_f,K_i}
f^{\sigma_f I_f}_{\kappa_f K_f}
f^{\sigma_i I_i}_{\kappa_i K_i}
\nonumber\\
&\quad \times
\langle \Phi_{\kappa_f} |
\hat{P}^{I_f}_{K_f}
\hat{O}_{L}
\hat{P}^{I_i}_{K_i}
| \Phi_{\kappa_i} \rangle .
\end{align}
By exploiting the properties of the projection operator, this expression reduces to integrals over Euler angles involving Wigner $D$ functions and intrinsic matrix elements of the form
$\langle \Phi_{\kappa_f} | \hat{O}_{L\mu} \hat{R}(\Omega) | \Phi_{\kappa_i} \rangle$,
which are evaluated numerically. 

The effective charges of
$e_{\pi}=1.5e$ for protons and $e_{\nu}=0.5e$ for neutrons are adopted, yielding a reliable description of $E2$ matrix elements calculated using the TPSM wave
functions~\cite{JS21}. The ability to compute complete sets of reduced $E2$ matrix elements enables stringent comparisons with modern multistep Coulomb-excitation measurements~\cite{nazira,rouoofe2}.

\section{RESULTS AND DISCUSSION}\label{Results}

The TPSM calculations for the six nuclides of $^{70}$Ge, $^{76,78,80,82}$Se and $^{100}$Mo
have been performed with the deformation parameters given in Table \ref{tab:parameters}. In the following subsections, we shall
discuss separately the results on excitation energies, $E2$ matrix elements and the derived shape invariants.

\subsection {Excitation energies}

TPSM results on excitation energies of some studied nuclides have already been published in our earlier works
\cite{JS21,Rouoof2024}. However, the focus has been primarily on the $\gamma$ bands and, in the present work, we have investigated all the
band structures for which $E2$ matrix elements are known. In particular, the results on excited $0^+$ bands, which were not studied
in our earlier works, will be discussed. In Fig.~\ref{E_core}, a detailed comparison of the TPSM calculated yrast, 
$\gamma$ and excited $0^+$ bands with the corresponding experimental bands are presented. The energies have been subtracted by
 a core contribution so that the differences can be visualized more clearly. 
 In Tab. \ref{Energy}, the energy values are also listed for detailed comparisons with observed energies and other
 model predictions. It is
 noted from Fig.~\ref{E_core} that yrast band,
which is known in all the cases up to quite high-spin, is reproduced quite well by the  TPSM approach. $\gamma$ band
is also known up to high-spin in some cases and it is evident from the figure that TPSM provides a good description
of this band structure as well.

\begin{table*}
\begin{longtable*}{p{2.4cm}p{2.4cm}p{2.4cm}p{2.4cm}p{2.4cm}p{2.4cm}p{2.4cm}p{0.9cm}}
\caption{  \label{tab:Pairing} Normal and reduced  neutron ($\Delta_{\nu}$) and proton ($\Delta_{\pi}$) pairing gaps for $^{70}$Ge, $^{76,78,80,82}$Se and $^{100}$Mo isotopes. }\\
 
\hline\hline
$\Delta$	&	$^{70}$Ge	&	$^{76}$Se  	&	 $^{78}$Se	&	$^{80}$Se	&	$^{82}$Se	&	$^{100}$Mo	\\
								\hline	\hline%
\endfirsthead													
													
\multicolumn{7}{c}{Normal pairing}\\													
\cline{1-7}													
													
$\Delta_{\nu}$	&	1.1987	&	0.9826	&	0.9672	&	1.0639	&	0.8089	&	1.1053	\\
$\Delta_{\pi}$	&	1.0937	&	1.0870	&	1.0627	&	1.0463	&	0.8681	&	0.9742	\\
													
\hline													
\multicolumn{7}{c}{Reduced pairing }\\													
\cline{1-7}													
$\Delta_{\nu}$	&	0.8584	&	0.4162	&	0.6903	&	0.5049	&	0.4575	&	0.6523	\\
$\Delta_{\pi}$	&	0.7044	&	0.7274	&	0.8972	&	0.7395	&	0.6861	&	0.4171	\\

\hline\hline

\end{longtable*}  
\end{table*}

As in our earlier publications \cite{nazira,rouoofe2}, we have performed a second set of TPSM
calculations with a reduced pairing such that the calculated band heads of the $0^+$ excited bands become close 
 to the known experimental energies. Table \ref{tab:Pairing} lists the pertaining gap parameters.
  The TPSM calculations with the pairing parameters as used
for the ground-state band gives too large energies for the excited $0^+$ bands.
By reducing the pair gaps, we are effectively taking into account the blocking effects that a proper self-consistent treatment 
 of the pair correlations would entail.
  The excited $0^+$ band is known up to $I$=8 in $^{70}$Ge and a few low-lying states are observed for other
nuclei as well. It is evident from Fig.~\ref{E_core} that TPSM reproduces these excited states reasonably well with the reduced pair gaps.

It has been shown in the framework of the collective models, based on the quadrupole degree of freedom, that energy staggering of 
the $\gamma$ band 
\begin{equation}
S(I)=(E(I)-2E(I-1)+E(I-2))/ E(2^+_1),
\end{equation}
is correlated with the softness 
of the triaxial deformation (see e.g., \cite{Frauendorf15a,Rouoof2024} ). 
The  extreme cases are the  Wilets-Jean limit of $\gamma$ independent potential,
the rigid $\gamma=30^\circ$ of triaxial rotor and that of the  axial rotor with the energy staggering, respectively, given by
 \begin{eqnarray}
 S_{WJ}(I)=\frac{1}{8}\mp\frac{1}{4}\left(I+\frac{9}{2}\right),\label{SWJ}\\
 S_{TR}(I)=\frac{1}{6}\pm\left(I-\frac{5}{2}\right),\label{STR}\\
 S_{AR}=\frac{1}{3}\label{SAR}.
 \end{eqnarray}  
Clearly, the properties of real nuclei lie between these extreme cases. A more subtle classification has been introduced in Ref. \cite{Rouoof2024} (page 4 ff.),
 which solved the Bohr Hamiltonian for a selection of characteristic potentials in the $\gamma$ degree of freedom. This work classified the potentials 
 in terms of the ground state and $\gamma$ band solutions as "axial" when $\gamma$ fluctuates about the prolate or oblate values and "triaxial"
 when $\gamma$ fluctuates about a finite value. Further  as "rigid" when the size of the fluctuations $\Delta\gamma \leq 16^\circ$, as "soft"
when $16^\circ \leq \Delta\gamma \leq 25^\circ$ and as 
"shallow" when  $\Delta\gamma \geq 25^\circ$ ( see the list of criteria on page 8 of Ref.  \cite{Rouoof2024}).

The energy head of the $\gamma$ band should be low  ($E(2_2)/E(4_1)$ close to or lower than one) for substantial triaxiality.
For triaxial $\gamma$ soft nuclei, the energy staggering 
parameter $S(I)$  has even-spin members below the odd-spin states and for triaxial
$\gamma$ rigid, the odd-spin members are favored in energy. The amplitude of the staggering is always much lower than the one of the 
limiting cases (Eqs. (\ref{SWJ}) and (\ref{STR})). 
In our earlier work  \cite{JS21,Rouoof2024,nazira,rouoofe2}, the staggering of
about twenty-three nuclei were studied and it was shown that only six nuclei depicted odd-spin members lower than
even-spin states and were categorized as triaxial $\gamma$ rigid.
However, it is noted that these energy criteria are only valid within the collective model and follow the
trends derived from the shape 
invariants, as discussed below. In the TPSM approach, on the other hand, this is not always true and is also evidenced in the experimental data 
(see the discussion of the Os and Pt isotopes in Ref. \cite{rouoofe2}). 

For $^{70}$Ge, the TPSM predicts  $S(I)$ for even-spin values,
initially, lower than the odd-spin ones. However, for the intermediate spin values the phase reverses. This change of phase 
is due to the admixing of the quasiparticle states into the vacuum configuration. There are too few experimental energies known  to verify this
prediction.

In the case of $^{76}$Se, data is known up to quite high spin and it is evident from Fig.~\ref{Stag} that TPSM  calculated
$S(I)$ agrees qualitatively with the experimental one.
Up to $I=14$, there is no clear staggering and above this spin value the TPSM predicts 
the  odd-spin values lower than the even-spin ones in accordance with the experimental data.   
For the three Se-isotopes,  $^{78,80,82}$Se the TPSM predicts 
 almost no staggering,  except for $^{82}$Se
in the high-spin region.  There is again no experimental
data for the $\gamma$ bands to verify these predictions. 
For $^{100}$Mo, the  even-spin-down pattern appears at $I=8$ in the TPSM calculations.

\subsection{ $E2$ matrix elements }

As in our earlier work, we have evaluated all $E2$ matrix elements up to $I=10$  for the studied six nuclides 
and are listed in the Tables \ref{E2T1} to \ref{E2T8}. The in-band matrix elements are also displayed in Figs.~\ref{E2-dia-y} to \ref{E22-ex0} and inter-band in Figs. \ref{E20-gy} to \ref{E22-gy}
in order to illustrate their angular-momentum  dependence.
The diagonal matrix elements along the yrast band, shown in Fig.~\ref{E2-dia-y}, are small. Their absolute values  are about 5-8 times smaller than
the $3_1\rightarrow 2_2$  values
 in Tab. \ref{E2T5}, which indicates $\gamma$ values are not far from $30^\circ$ (see Tab. 2 of Ref. \cite{Rouoof2024}). They 
depict slightly positive values for $^{70}$Ge, and slightly negative values for all other isotopes, which implies a tendency toward the
oblate shape in $^{70}$Ge and toward the prolate shape for the other isotopes.  In $^{70}$Ge,
the $E2$ matrix element becomes negative for $I=8$ and for $^{76}$Se, they
become positive for $I=10$. TPSM predicts a shape change for these two spin states and the experimental data is needed
to confirm this predicted shape change.

The yrast $I$ to $(I-2)$ transitions are displayed in Fig.~\ref{E22-in-y}, which have tendency to increase with spin. The calculated transition matrix elements
are in good agreement with the experimental values. For the $\gamma$ band, the diagonal and non-diagonal in-band transitions are depicted in
Figs.~\ref{E2-dia-g} and \ref{E22-in-g}, respectively. The diagonal matrix elements in $^{70}$Ge are all negative, except for
the $4^+$ state. For all other nuclides, the diagonal matrix elements of the $2^+$ state are positive and all other
are negative. The measured values are known for a few cases, which  
are in good agreement with TPSM calculations. The $I\rightarrow (I-2)$ in-band transitions of the $\gamma$ band are almost  constant
 with spin. The few measured values are in reasonable agreement with the TPSM  values. 

\subsection{ Shape invariants }

The shape invariant quantities are presented in Figs.~\ref{f1:sigmaQ2} to \ref{f6:sigmaQ2} for the six nuclides studied. These quantities
have been calculated by means of the GOSIA code \cite{GOSIA2012}   and using the TPSM calculated $E2$ matrix elements given Tables \ref{E2T1} to \ref{E2T8}. The experimental
shape invariant values provided in the figures are those available in the published articles
\cite{cmks_jst1_70ge,KAVKA1995177,Hayakawa2003,Wrzosek064305}. In addition, Fig.~\ref{f7:delta_bands} 
summarizes the TPSM results by depicting for each nuclide an estimate of the range within which the  fluctuating
triaxiality parameter $\delta$ is confined. The lower  
limit is $\delta_\textrm{min}=\arccos\left[\langle\cos 3\delta\rangle+\sigma(\textrm {cos}3\delta)\right]/3$, 
the upper limit is $\delta_\textrm{max}=\arccos\left[\langle\cos 3\delta\rangle-\sigma(\textrm {cos}3\delta)\right]/3$ and
the middle line displays the centroid  $\delta_\textrm{mean}=\arccos\left[\langle\cos 3\delta\right]/3$. Considering that 
the fluctuations obey a normal distribution of width $\sigma$, then the probability of $\delta$ 
to be within the lower and upper bounds is 68\% as also discussed in Ref. \cite{Henderson054306}.

In the case of $^{70}$Ge, shown in Fig.~\ref{f1:sigmaQ2}, $\langle Q^2\rangle$ is about 0.29 $\textrm{e}^2\textrm{b}^2$ for the yrast configuration, and also has a similar value for the
$\gamma$ band. This value corresponds to the deformation, $\beta\approx 0.289$. The fluctuations of $\langle Q^2\rangle$,
designated by $\sigma (Q^2)$, are small.
This means that the admixture of quasiparticle configurations does not generate substantial fluctuations in the $\beta$ degree of freedom. The small dispersion values
for other nuclides indicate a similar stability of the $\beta$ deformation.

The value of $\langle\cos 3\delta\rangle$ for the yrast states is about -0.18, which corresponds to $\delta\approx33^o$.
The slight tendency toward oblate shape is already indicated by the diagonal matrix elements. 
 For the $\gamma$ band, $\langle\cos 3\delta\rangle$ is positive for  most of the states (except for $I=4$), which indicates a tendency toward the oblate shape 
 as the diagonal matrix elements depict. The fluctuations in $\cos 3\delta$ are
quite significant for most of the spin states along both yrast and $\gamma$ bands.
The wide band of $\delta$ in Fig.  \ref{f7:delta_bands}, signifies  a
$\gamma$ soft character and is  consistent with the even-$I$-down staggering of the $\gamma$ band energies in Fig. \ref{Stag}.

The shape invariants  for $^{76}$Se are depicted in Fig.~\ref{f2:sigmaQ2} and
 for both yrast and the $\gamma$ bands, $\langle Q^2\rangle$ shows a slight drop with increasing spin.
 The triaxiality parameter $\langle\cos 3\delta\rangle$ is small and its yrast value of  0.01  for $I=0$ ($\delta=29.7^\circ$) increases with $I$, which
 shifts $\delta$ 
toward  the prolate shape as seen Fig. \ref{f7:delta_bands}. The slight preference of the prolate sector is consistent
  with the diagonal matrix elements in Fig. \ref{E2-dia-y}. The triaxiality parameter $\delta$ of the $\gamma$ band
  slightly decreases with $I$ as well.
  The fluctuations  $\sigma(\cos 3\delta)$ of both bands are quite small, but less regular for the $\gamma$ band.
  The narrow limits in Fig. \ref{f7:delta_bands} classify  $^{76}$Se as a rigid triaxial system. 
  This is at variance with the small irregular values of $S(I)$ in Fig. \ref{Stag}, which  are expected for the transition  between $\gamma$ soft and
  $\gamma$ rigid regions of the collective model.
  
 For $^{78}$Se in Fig.~\ref{f3:sigmaQ2}, TPSM values $\langle Q^2\rangle$ for both yrast and $\gamma$ bands are almost constant and the 
fluctuation in $\langle Q^2\rangle$ are quite small.
The centroid $\langle\cos 3\delta\rangle$ is not  small and its yrast value  for $I=0$ of  0.383  corresponds to $\delta=22.5^\circ$, well
located in the prolate sector $\delta<30^\circ$, which is consistent
  with the diagonal matrix elements depicted in Fig. \ref{E2-dia-y}. 
 The fluctuation $\sigma(\cos 3\delta)$ increases from 0 at $I=0$ to  0.5 at $I=8$.
  Correspondingly,  in Fig. \ref{f7:delta_bands}, $\delta(I)$ increases
  to about $30^\circ$ at $I=6$ and the fluctuation range strongly widens. 
 In the collective model picture, this is interpreted as a transition from a stable triaxial shape with $\gamma\approx 22^\circ$
 to a $\gamma$ soft shape centered at $30^\circ$.
   For the $\gamma$ band, the value of $\cos 3\delta\approx 0.5$ and its dispersion of $\sigma(\cos 3\delta)\sim 0.4 $ indicate large fluctuation around the moderate-triaxial shape 
   of $\delta=20^\circ$. This is consistent with the even-$I$-low pattern of $S(I)$ for $I \leq 7$ 
   in Fig.  \ref{Stag}, but not with its suppression at higher $I$.
 
 For $^{80}$Se, the shape invariants are shown in Fig.~\ref{f4:sigmaQ2} with $\langle Q^2\rangle$ being almost constant for the yrast states
and slight drop for the $\gamma$ band. The small dispersions $\langle Q^2\rangle$ indicate a stable $\beta$ value for both the bands. The value of
$\langle\cos 3\delta\rangle=0.0505$ at $I=0$ corresponds to $\delta=29^\circ$ that slightly decreases with $I$.
The dispersion $\sigma(\cos 3\delta)$ increases with $I$ from a small to a moderate value, which is close to maximal triaxiality.
 The $\gamma$ band has slightly positive values 
corresponding to $\delta\sim 25^\circ$,  which implies lower triaxiality. The fluctuations of $\langle\cos 3\delta\rangle$ decrease with $I$ from a moderate to a small value. In the context of the collective models, Fig. \ref{f7:delta_bands} locates
both bands in between $\gamma$ rigid and soft, which is consistent with the diagonal quadrupole moments, shown in
Figs.  \ref{E2-dia-y} and  \ref{E2-dia-g}. The small experimental values of  $\sigma(\cos 3\delta)$ seems to indicate
a more rigid triaxial shape as compared to the TPSM predictions. It is to be noted that staggering pattern in Fig. \ref{Stag} does
not correlate with the shape invariants.

The TPSM values for $^{82}$Se, shown in Fig.~\ref{f5:sigmaQ2}, are similar to those of $^{80}$Se, where the experimental values point to a lower $\beta$ value.
The  mean value for the yrast band  of $\langle\cos 3\delta\rangle= 0.407$ at $I=0$ indicates moderate triaxiality of $\delta=22.0^\circ$, which decreases with $I$. The dispersion 
 $\sigma(\cos 3\delta)$ is small. The moderate triaxiality predicted by the TPSM is at variance with the experiment, which points to near-maximum
triaxiality, similar to $^{80}$Se. The $\delta$ values are consistent with diagonal matrix elements in Fig. \ref{E2-dia-y}, which indicate
near-maximal triaxiality in 
 agreement with the experiment. For the $\gamma$ band, the TPSM values of  $\langle\cos 3\delta\rangle$ predict 
 a change from moderate triaxiality of $\delta=20^\circ$ at $I=2$ to maximal triaxiality.  The dispersion 
 $\sigma(\cos 3\delta)$ is small. 
 Experimental values for $\delta$ are not available. 
 The diagonal matrix elements in Fig. \ref{E2-dia-g} agree well with the experimental data.

The shape invariants for $^{100}$Mo are displayed in Fig.~\ref{f6:sigmaQ2} with $\langle Q^2\rangle$ being nearly constant. 
For the $\gamma$ band, $\langle Q^2\rangle$ decreases with $I$.
The average triaxiality  $\langle\cos 3\delta\rangle=-0.141$ for the $I$=0  yrast state is  located at $32.7^\circ$, which is  close to  maximal triaxiality  on the oblate side. 
The experimental value  corresponds to the maximum $\delta=30^\circ$ within the error limits. 
For the $I>0$ yrast states, $\langle\cos 3\delta\rangle$  remains small.  The small diagonal matrix elements in Fig. \ref{E2-dia-y} are consistent with 
near-maximum average triaxiality with a slight tendency toward the oblate side with increasing $I$. The TPSM values of $\langle\cos 3\delta\rangle$  for the
$\gamma$ band are first slightly positive, i.e., somewhat in the prolate sector, but approach zero with $I$.  
The positive value of the diagonal matrix elements in Fig. \ref{E2-dia-y}  
also point to   a slight prolate preference that decreases with $I$.   
The dispersions $\sigma(\cos 3\delta)$ are small  for both bands.  The corresponding narrow fluctuation ranges 
in Fig. \ref{f7:delta_bands} classify $^{100}$Mo as $\gamma$ rigid, which is at variance
with the even-$I$-down pattern of $S(I)$ seen in Fig. \ref{Stag}. It is noted that the experimental
values support  this deviation of the TPSM values
from the correlation between energy staggering and $\gamma$ softness indicated by the collective model.

\section{Summary and conclusions}\label{Sum_Con}

The present manuscript is a continuation of our earlier works \cite{nazira,rouoofe2} to calculate  the $E2$
matrix elements and the corresponding shape invariant quantities by means of the microscopic TPSM approach, and compare them  with the
available experimental data. We have
investigated six nuclides of $:$ $^{70}$Ge, $^{76,78,80,82}$Se and  $^{100}$Mo for which  reasonably extended data sets  
from  COULEX experiments are available. For completeness, we have also calculated excitation energies and other properties of
the studied nuclides in order to validate the predictions of the TPSM approach.

It has been demonstrated that energies  of the observed band structures are well reproduced by the TPSM calculations
for all the six nuclides. It is noted that within the collective model, the staggering parameter of the $\gamma$ band energies is
sensitive to the nature
of the $\gamma$ deformation $:$ the even-$I$-low pattern  signifies  that the mode is 
$\gamma$ soft, whereas the odd-$I$-down pattern signifies a more rigid triaxial shape. 
It has been shown  that TPSM values show a complex $I$-dependence.  For the nuclides of $^{70}$Ge and $^{76}$Se, the staggering parameter changes phase 
from even-$I$-low to odd-$I$-low, whereas for $^{100}$Mo the  opposite phase is predicted. For the other nuclides, the staggering  
becomes quite small for  certain intervals of $I$. 
In all cases, except for $^{76}$Se,  the observed energies of the $\gamma$ band are available only
for low-spin states and the predicted complex $I$-dependence cannot be verified.

The main objective of the work has been to evaluate the $E2$ matrix elements. For each nucleus we have
evaluated 103 matrix elements and have been tabulated in Tables \ref{E2T1} to \ref{E2T8} for furture experimental and theoretical comparisons. 
The calculated matrix elements are in good agreement with the observed values from the COULEX data, wherever available. 
Using these $E2$ matrix elements, we have deduced the shape invariants with the Kumar-Cline sum rules. In most
of the cases, it has been shown that shape is $\gamma$ soft. For $^{76}$Se and $^{100}$Mo, the sum rules indicate 
$\gamma$ rigidity, however, a complex 
staggering pattern  of the $\gamma$ band energies is predicted by the TPSM calculations. It would be quite interesting to 
measure the $\gamma$ band energies of the nuclides to validate the TPSM predictions.

The direct inference from the $E2$ matrix elements for $^{76}$Se and $^{100}$Mo 
is in conflict with the phenomenological collective models having the 
quadrupole degree of freedom, which predict  an odd-$I$-down pattern for $\gamma$ rigid nuclei. As discussed in 
our previous works \cite{nazira,Rouoof2024,rouoofe2}, the staggering appears due to the interaction of 
the $\gamma$ band with ground band and nearby even-$I$ rotational bands. In the collective model  this is
the vibrational excitation in the $\gamma$ degree of freedom, which is sensitive to the stiffness of the potential.
In the microscopic TPSM approach, there are several  two-quasiparticle excitations that generate a more complex pattern.
The results of the present study provide examples on the limitations of the collective model.

We would like to mention that large sets of COULEX data are also available for $^{106,108,110}$Pd \cite{Svensson:1995lek,DC86} and $^{114}$Cd \cite{FAHLANDER1988327}.
These systems have near spherical or weakly deformed shapes
and TPSM aproach is not directly applicable to them since it employs a single deformed Slater determinant as the intrinsic configuration. The extended TPSM
approach that considers a set of deformed configurations in the framework of generator coordinate method (GCM) is needed to investigate the properties
of weakly deformed nuclei where shapes are not stable. This generalized TPSM approach is presently being developed and its application to COULEX data of near spherical
nuclides shall be presented in a forthcoming publication.

\section*{Acknowledgements}
The authors acknowledge the Board of Research in Nuclear Sciences (BRNS), Department of Atomic Energy (DAE), Government of India, for financial assistance through Project No. 58/14/08/2025-BRNS/290.

\begin{longtable*}{p{2.65cm}p{2.65cm}p{2.65cm}p{2.65cm}p{2.65cm}p{2.65cm}c}
\caption{TPSM and experimental energies (MeV) of the yrast, $\gamma$, and $0_2^+$ bands of $^{70}$Ge, $^{76,78,80,82}$Se and $^{100}$Mo isotopes. Available  experimental values are
listed in parenthesis below the TPSM values.}\\
 
\hline\hline
Spin	&	$^{70}$Ge	&	$^{76}$Se  	&	 $^{78}$Se	&	$^{80}$Se	&	$^{82}$Se	&	$^{100}$Mo		\\
								\hline							
\endfirsthead															
															
\multicolumn{7}{c}{yrast band}\\															
\cline{1-7}															
0	&	0	&	0	&	0	&	0	&	0	&	0		\\
	&	(0)	&	(0)	&	(0)	&	(0)	&	(0)	&	(0)		\\
2	&	0.838	&	0.559	&	0.595	&	0.652	&	0.662	&	0.472		\\
	&	(1.040)	&	(0.559)	&	(0.613)	&	(0.666)	&	(0.654)	&	(0.535)		\\
4	&	1.948	&	1.355	&	1.497	&	1.624	&	1.799	&	1.170		\\
	&	(2.153)	&	(1.331)	&	(1.502)	&	(1.702)	&	(1.735)	&	(1.136)		\\
6	&	3.181	&	2.259	&	2.479	&	2.736	&	3.007	&	2.026		\\
	&	(3.297)	&	(2.262)	&	(2.546)	&	(2.895)	&	(3.144)	&	(1.847)		\\
8	&	4.231	&	3.130	&	3.522	&	3.651	&	3.529	&	2.699		\\
	&	(4.204)	&	(3.269)	&	(3.585)	&	(3.635)	&	(3.517)	&	(2.627)		\\
10	&	5.409	&	4.134	&	4.519	&	4.813	&	5.465	&	3.200		\\
	&	(5.243)	&	(4.299)	&	(4.625)	&	(4.846)	&	(5.457)	&	(3.367)		\\
\hline															
															
\multicolumn{7}{c}{$\gamma$ band}\\															
\cline{1-7}															
															
2	&	1.704	&	1.152	&	1.305	&	1.385	&	1.774	&	0.967		\\
	&	(1.708)	&	(1.216)	&	(1.318)	&	(1.449)	&	(1.731)	&	(1.064)		\\
3	&	2.483	&	1.699	&	1.872	&	1.974	&	2.385	&	1.435		\\
	&	(2.451)	&	(1.689)	&	(1.853)	&	    		&	(2.550)	&	(1.607)		\\
4	&	2.802	&	2.054	&	2.210	&	2.512	&	2.901	&	2.186		\\
	&	(2.806)	&	(2.026)	&	(2.190)	&		&		&	(2.310)		\\
5	&	3.547	&	2.540	&	2.841	&	3.078	&	3.526	&	2.574		\\
	&	(3.669)	&	(2.486)	&	(2.735)	&		&		&			\\
6	&	3.756	&	2.925	&	3.233	&	3.571	&	3.957	&	2.922		\\
	&	(3.753)	&	(2.976)	&	(3.140)	&		&		&			\\
7	&	3.970	&	3.377	&	3.763	&	4.108	&	4.583	&	3.178		\\
	&		&	(3.432)	&	(3.704)	&		&		&			\\
8	&	4.987	&	3.984	&	4.389	&	4.198	&	4.616	&	3.198		\\
	&	(4.820)	&	(3.853)	&	(3.830)	&		&		&			\\
9	&	5.233	&	4.602	&	5.094	&	4.881	&	4.867	&	3.576		\\
	&		&	(4.405)	&		&		&		&			\\
10	&	5.939	&	4.748	&	5.769	&	5.463	&	6.308	&	3.667		\\
	&		&	(4.687)	&		&		&		&			\\
\hline															
\multicolumn{7}{c}{${0_2^+} $ band}\\															
\cline{1-7}															
0	&	1.132	&	1.372	&	1.248	&	1.46	&	1.458	&	0.696		\\
	&	(1.216)	&	(1.122)	&	(1.478)	&	(1.479)	&	(1.410)	&	(0.695)		\\
2	&	2.153	&	1.701	&	1.6	&	1.877	&	2.122	&	1.506		\\
	&	(2.157)	&	(1.787)	&		&		&		&	(1.639)		\\
4	&	3.089	&	2.309	&	2.251	&	2.627	&	2.939	&	2.05		\\
	&	(3.059)	&		&		&		&		&			\\
6	&	4.235	&	3.298	&	3.072	&	3.697	&	4.043	&	2.44		\\
	&	(4.104)	&		&		&		&		&			\\
8	&	5.447	&	4.478	&	4.358	&	5.089	&	5.236	&	3.24		\\
	&	(5.436)	&		&		&		&		&			\\
10	&	7.040	&	5.939	&	5.287	&	6.782	&	6.283	&	4.067		\\

\hline\hline
  
   \label{Energy}

\end{longtable*}  


\begin{longtable*}{p{2.52cm}p{2.52cm}p{2.52cm}p{2.52cm}p{2.52cm}p{2.52cm}c}
\caption{Reduced TPSM  $E2$ diagonal  matrix elements (in eb units) in $^{70}$Ge, $^{76,78,80,82}$Se and $^{100}$Mo. Available  experimental matrix elements are
listed in parenthesis below the TPSM values, where the errors are quoted in square brackets.}\\
 
\hline\hline
$I_i \rightarrow I_f$	&	$^{70}$Ge	&	$^{76}$Se  	&	 $^{78}$Se	&	$^{80}$Se	&	$^{82}$Se	&	$^{100}$Mo	\\
								\hline	\hline%
\endfirsthead													
													
\multicolumn{7}{c}{$I_{yrast}  \rightarrow  I_{yrast}$}\\													
\cline{1-7}													
$2_1\rightarrow 2_1$	&	0.181	&	-0.315	&	-0.49	&	-0.37	&	-0.428	&	-0.13	\\
	&	(0.22 [{$^{+4}_{-3}$}])	&	(-0.45 [$^{+0.07}_{-0.07}$])	&	(-0.27 [$^{+0.09}_{-0.09}$])	&	(-0.26[$^{+0.04}_{-0.03}$])	&	(-0.30[$^{+0.04}_{-0.03}$])	&	(-0.33 [$^{+0.10}_{-0.10}$])	\\
$4_1\rightarrow 4_1$	&	0.18	&	-0.189	&	-0.81	&	-0.761	&	-0.655	&	-0.57	\\
	&	(0.14 [{$^{+5}_{-6}$}])	&	(-0.36 [{$^{+0.24}_{-0.14}$}])	&	(-0.90 [{$^{+0.20}_{-0.20}$}])	&	(-0.85 [$^{+0.11}_{-0.06}$])	&	(-0.76 [$^{+0.07}_{-0.08}$])	&	(-0.35 [$^{+0.18}_{-0.18}$])	\\
$6_1\rightarrow 6_1$	&	0.051	&	-0.203	&	-0.097	&	-0.205	&	-0.701	&	-0.73	\\
	&	(0.1 [{$^{+2}_{-4}$}])	&		&		&		&		&		\\
$8_1\rightarrow 8_1$	&	-0.966	&	-0.126	&	-0.343	&	-0.446	&	-1.248	&	-0.78	\\
$10_1\rightarrow 10_1$	&	-0.309	&	0.843	&	-0.389	&	-0.374	&	-1.436	&	-0.74	\\
													
\hline													
													
\multicolumn{7}{c}{$I_{\gamma} \rightarrow I_{\gamma}$}\\													
\cline{1-7}													
													
$2_2\rightarrow 2_2$	&	-0.318	&	0.152	&	0.38	&	0.365	&	0.395	&	0.904	\\
	&	(-0.44 [{$^{+4}_{-3}$}])	&	(0.24 [{$^{+0.06}_{-0.08}$}])	&	(0.23 [{$^{+0.12}_{-0.12}$}])	&	(0.53 [$^{+0.03}_{-0.03}$])	&	(0.45 [{$^{+0.04}_{-0.05}$}])	&	(1.20 [$^{+0.10}_{-0.08}$])	\\
$3_1\rightarrow 3_1$	&	0.019	&	0.002	&	0.004	&	0.005	&	-0.002	&	0.009	\\
$4_2\rightarrow 4_2$	&	0.566	&	-0.365	&	-0.37	&	-0.41	&	-0.384	&	-0.538	\\
	&	(0.7 [{$^{+3}_{-2}$}])	&		&		&		&	(-0.40 [{$^{+0.13}_{-0.30}$}])	&		\\
$5_1\rightarrow 5_1$	&	-0.264	&	-0.239	&	-0.204	&	-0.235	&	-0.269	&	-0.258	\\
$6_2\rightarrow 6_2$	&	-0.44	&	-0.523	&	-0.466	&	-0.503	&	-0.681	&	-0.696	\\
$7_1\rightarrow 7_1$	&	-0.603	&	-0.331	&	-0.243	&	-0.299	&	-0.477	&	-0.335	\\
$8_2\rightarrow 8_2$	&	-0.644	&	-0.451	&	-0.268	&	-0.337	&	-0.249	&	-0.418	\\
$9_1\rightarrow 9_1$	&	-0.853	&	-0.254	&	-0.413	&	-0.461	&	-1.131	&	-0.528	\\
$10_2\rightarrow 10_2$	&	-1.06	&	-0.419	&	-0.301	&	-0.347	&	-1.031	&	-0.362	\\
													
\hline													
\multicolumn{7}{c}{$I_{0_2^+} \rightarrow I_{0_2^+}$}\\													
\cline{1-7}													
$2_3\rightarrow 2_3$	&	0.54	&	-0.264	&	0.431	&	-0.29	&	-0.346	&	-0.595	\\
	&	(0.49 [{$^{+27}_{-8}$}])	&		&		&		&		&	(-0.24 [$^{+0.12}_{-0.07}$])	\\
$4_3\rightarrow 4_3$	&	0.14	&	-0.129	&	0.291	&	-0.568	&	-0.633	&	-0.867	\\
$6_3\rightarrow 6_3$	&	0.04	&	-0.085	&	-0.351	&	-0.193	&	-0.264	&	-1.009	\\
$8_3\rightarrow 8_3$	&	-0.473	&	-0.074	&	-0.417	&	-0.099	&	-0.249	&	-0.337	\\
$10_3\rightarrow 10_3$	&	-1.426	&	-0.068	&	-0.792	&	-0.067	&	-0.029	&	-0.277	\\
		
\hline\hline
  
   \label{E2T1}

\end{longtable*}  

\begin{longtable*}{p{2.52cm}p{2.52cm}p{2.52cm}p{2.52cm}p{2.52cm}p{2.52cm}c}
\caption{Reduced TPSM $E2$ inter-band matrix elements (in eb units) for  $I\rightarrow I $ transitions in $^{70}$Ge, $^{76,78,80,82}$Se and $^{100}$Mo.
Available  experimental matrix elements are
listed in parenthesis below the TPSM values, where the errors are quoted in square brackets.
}\\

\hline\hline
$I_i \rightarrow I_f$	&	$^{70}$Ge	&	$^{76}$Se  	&	 $^{78}$Se	&	$^{80}$Se	&	$^{82}$Se	&	$^{100}$Mo	\\
								\hline	\hline%
\endfirsthead													
													
\multicolumn{7}{c}{$I_{\gamma}\rightarrow  I_{Yrast}$ }\\													
\cline{1-7}													

$2_2\rightarrow 2_1$	&	0.651	&	0.562	&	0.517	&	0.576	&	0.153	&	0.804	\\
	&	(0.416 [$^{+13}_{-9}$])	&	(0.667 [$^{+0.036}_{-0.036}$])	&	(0.45 [$^{+0.04}_{-0.04}$])	&	(0.379 [$_{-0.022}^{+0.020}$ ])	&	(0.19 [$^{+0.015}_{-0.022}$])	&	(0.94 [$^{+0.02}_{-0.02}$])	\\
$4_2\rightarrow 4_1$	&	0.318	&	0.261	&	0.552	&	0.391	&	0.291	&	0.987	\\
	&	(0.38 [$^{+5}_{-3}$])	&	(0.10 [$^{+0.05}_{-0.04}$])	&		&		&	(0.28 [$^{+0.05}_{-0.04}$])	&	(0.99 [$^{+0.05}_{-0.05}$])	\\
$6_2\rightarrow 6_1$	&	0.105	&	0.265	&	0.454	&	0.264	&	0.335	&	1.95	\\
$8_2\rightarrow 8_1$	&	0.679	&	0.061	&	0.534	&	0.262	&	0.437	&	0.295	\\
$10_2\rightarrow 10_1$	&	0.528	&	0.07	&	0.389	&	0.237	&	0.801	&	0.406	\\
	&		&		&		&		&		&		\\
													
\hline													
\multicolumn{7}{c}{$I_{0_2^+} \rightarrow  I_{Yrast}$ }\\													
\cline{1-7}													
$2_3\rightarrow 2_1$	&	0.47	&	0.027	&	0.03	&	0.082	&	0.132	&	-0.015	\\
	&	(0.276 [$^{+5}_{-7}$])	&	(0.07 [$^{+0.01}_{-0.01}$])	&		&	(0.10 [$^{+0.02}_{-0.03}$])	&	(0.10 [$^{+0.05}_{-0.07}$])	&	(-0.07 [$^{+0.007}_{-0.006}$])	\\
$4_3\rightarrow 4_1$	&	0.026	&	0.023	&	0.021	&	0.04	&	0.217	&	0.038	\\
$6_3\rightarrow 6_1$	&	-0.132	&	0.025	&	-0.002	&	-0.079	&	0.025	&	0.042	\\
$8_3\rightarrow 8_1$	&	-0.016	&	-0.09	&	0.043	&	-0.091	&	0.624	&	0.14	\\
$10_3\rightarrow 10_1$	&	0.036	&	-0.0007	&	0.175	&	-0.025	&	0.141	&	0.031	\\
													
\hline													
\multicolumn{7}{c}{$I_{0_2^+} \rightarrow I_{\gamma}$ }\\													
\cline{1-7}													
$2_3\rightarrow 2_2$	&	0.502	&	0.3	&	-0.006	&	0.112	&	0.074	&	0.053	\\
	&	(0.41 [$^{+3}_{-4}$])	&	(${\pm 0.66}$[$^{+0.21}_{-0.14}$])	&		&	($\pm 0.07$[$^{+0.09}_{-0.10}$])	&	(0.01[$^{+0.09}_{-0.17}$])	&	(0.40 [$^{+0.15}_{-0.13}$])	\\
$4_3\rightarrow 4_2$	&	-0.007	&	-0.05	&	0.009	&	0.322	&	0.048	&	-0.265	\\
$6_3\rightarrow 6_2$	&	0.163	&	-0.215	&	-0.111	&	-0.147	&	0.006	&	-0.019	\\
$8_3\rightarrow 8_2$	&	-0.159	&	-0.048	&	-0.048	&	-0.081	&	0.345	&	0.175	\\
$10_3\rightarrow 10_2$	&	-0.05	&	-0.011	&	0.026	&	0.018	&	0.141	&	0.056	\\

\hline\hline																

  \label{E2T2}
\end{longtable*}

\begin{longtable*}{p{2.52cm}p{2.52cm}p{2.52cm}p{2.52cm}p{2.52cm}p{2.52cm}c}
\caption{Reduced TPSM $E2$ in-band matrix elements (in eb units) for transitions $I_i \rightarrow (I-2)_f$ in $^{70}$Ge,$^{76,78,80,86}$Se and $^{100}$Mo. 
Available  experimental matrix elements are listed in parenthesis below the TPSM values, where the errors are quoted in square brackets.
}\\

 \hline\hline
$I_i \rightarrow I_f$	&	$^{70}$Ge	&	$^{76}$Se	&	 $^{78}$Se	&	$^{80}$Se	&	$^{82}$Se	&	$^{100}$Mo	\\
													
\hline\hline
\endfirsthead													
													
\multicolumn{7}{c}{$I_{yrast} \rightarrow (I-2)_{yrast}$}\\													
\cline{1-7}													
$2_1\rightarrow 0_1$	&	0.561	&	0.484	&	0.651	&	0.513	&	0.501	&	0.681	\\
	&	(0.422 [4])	&	(0.647 [$^{+0.033}_{-0.033}$])	&	(0.57 [$^{+0.04}_{-0.04}$])	&	(0.486 [$^{+0.028}_{-0.025}$])	&	(0.423 [$^{+0.022}_{-0.022}$])	&	(0.68 [$^{+0.01}_{-0.01}$])	\\
$4_1\rightarrow 2_1$	&	0.671	&	1.056	&	1.058	&	0.622	&	0.835	&	1.18	\\
	&	(0.68 [$^{+5}_{-1}$])	&	(1.14 [$^{+0.06}_{-0.06}$])	&	(0.81 [$^{+0.06}_{-0.06}$])	&	(0.82 [$^{+0.04}_{-0.04}$])	&	(0.63 [$^{+0.03}_{-0.03}$])	&	(1.33 [$^{+0.03}_{-0.02}$])	\\
$6_1\rightarrow 4_1$	&	0.795	&	1.584	&	1.37	&	0.662	&	1.082	&	1.854	\\
	&	(0.75 [$^{+6}_{-8}$])	&	(1.53 [$^{+0.16}_{-0.19}$])	&		&		&		&	(1.83 [$^{+0.06}_{-0.06}$])	\\
$8_1\rightarrow 6_1$	&	0.962	&	1.596	&	1.384	&	0.843	&	1.296	&	1.941	\\
	&		&	(1.64 [$^{+0.22}_{-0.15}$])	&		&		&		&		\\
$10_1\rightarrow 8_1$	&	1.219	&	1.665	&	1.771	&	0.965	&	1.451	&	1.974	\\
\hline													
\multicolumn{7}{c}{$I_{\gamma} \rightarrow (I-2)_{\gamma}$ }\\													
\cline{1-7}													
$4_2\rightarrow 2_2$	&	0.546	&	0.732	&	0.658	&	0.361	&	0.607	&	0.982	\\
	&	(0.44 [$^{+3}_{-2}$])	&	(0.92 [$^{+0.13}_{-0.21}$])	&		&	(0.67 [$_{-0.18}^{+0.08}$])	&	(0.71 [$^{+0.03}_{-0.09}$])	&	(1.02 [$^{+0.04}_{-0.03}$])	\\
$5_1\rightarrow 3_1$	&	0.692	&	0.822	&	0.953	&	0.475	&	0.835	&	0.616	\\
$6_2\rightarrow 4_2$	&	0.293	&	0.947	&	0.687	&	0.288	&	1.054	&	0.46	\\
	&		&	(0.91 [$^{+0.28}_{-0.14}$])	&		&		&		&		\\
$7_1\rightarrow 5_1$	&	0.757	&	0.453	&	1.277	&	0.413	&	0.176	&	0.669	\\
$8_2\rightarrow 6_2$	&	0.568	&	0.098	&	1.144	&	0.399	&	0.705	&	0.039	\\
$9_1\rightarrow 7_1$	&	0.993	&	0.543	&	1.31	&	0.313	&	0.159	&	0.478	\\
$10_2\rightarrow 8_2$	&	0.998	&	0.594	&	1.42	&	0.615	&	0.784	&	0.361	\\
													
													
\hline													
\multicolumn{7}{c}{$I_{0_2^+}  \rightarrow (I-2)_{0_2^+}$}\\													
\cline{1-7}													
$2_3\rightarrow 0_2$	&	-0.552	&	0.42	&	0.433	&	0.382	&	0.133	&	0.668	\\
	&	(-0.349 [$^{+8}_{-9}$])	&	(0.59 [$^{+0.30}_{-0.74}$])	&		&	(0.23 [$^{+0.05}_{-0.16}$])	&	(0.23 [$^{+0.07}_{-0.35}$])	&	(0.506 [$^{+0.008}_{-0.006}$])	\\
$4_3\rightarrow 2_3$	&	0.838	&	0.517	&	0.528	&	0.177	&	0.858	&	-0.016	\\
$6_3\rightarrow 4_3$	&	0.451	&	0.159	&	0.395	&	0.228	&	0.154	&	0.369	\\
$8_3\rightarrow 6_3$	&	-0.764	&	-0.121	&	0.114	&	-0.068	&	0.029	&	-0.232	\\
$10_3\rightarrow 8_3$	&	-0.943	&	0.536	&	-0.066	&	0.525	&	0.833	&	0.653	\\

 \hline\hline
  \label{E2T3}
\end{longtable*}

\begin{longtable*}{p{2.52cm}p{2.52cm}p{2.52cm}p{2.52cm}p{2.52cm}p{2.52cm}c}
\caption{Reduced TPSM $E2$ inter-band matrix elements (in eb units) for transitions $I_i \rightarrow (I-2)_f$ in $^{70}$Ge, ,$^{76,78,80,82}$Se and $^{100}$Mo. 
Available  experimental matrix elements are
listed in parenthesis below the TPSM values, where the errors are quoted in square brackets.
}\\

\hline\hline
$I_i \rightarrow I_f$	&	$^{70}$Ge	&	$^{76}$Se	&	 $^{78}$Se	&	$^{80}$Se	&	$^{82}$Se	&	$^{100}$Mo	\\
													
\hline\hline
\endfirsthead													
													
\multicolumn{7}{c}{$I_{\gamma} \rightarrow (I-2)_{Yrast}$ }\\													
\cline{1-7}													
$2_2\rightarrow 0_1$	&	-0.021	&	0.061	&	0.035	&	0.144	&	0.133	&	0.131	\\
	&	(-0.0526 [8])	&	(0.112 [$^{+0.006}_{-0.006}$])	&	(0.08[$^{+0.01}_{-0.01}$])	&	(0.106 [$^{+0.006}_{-0.006}$])	&	(0.120 [$^{+0.006}_{-0.006}$])	&	(0.103 [$^{+0.002}_{-0.001}$])	\\
$4_2\rightarrow 2_1$	&	0.204	&	0.102	&	0.12	&	0.076	&	0.124	&	0.061	\\
	&		&	(0.12 [$^{+0.07}_{-0.09}$])	&		&	($\pm 0.01$[$^{+0.13}_{-0.06}$])	&	(0.09 [{$^{+0.01}_{-0.01}$}])	&	(0.063 [$^{+0.025}_{-0.012}$])	\\
$6_2\rightarrow 4_1$	&	0.761	&	0.163	&	0.179	&	0.145	&	0.399	&	0.125	\\
$8_2\rightarrow 6_1$	&	-0.653	&	0.134	&	0.331	&	0.438	&	-1.133	&	-0.362	\\
$10_2\rightarrow 8_1$	&	-0.267	&	-0.01	&	0.384	&	0.138	&	0.519	&	0.563	\\
													
\hline													
\multicolumn{7}{c}{$I_{0_2^+} \rightarrow (I-2)_{Yrast}$ }\\													
\cline{1-7}													
$2_3\rightarrow 0_1$	&	0.054	&	0.028	&	-0.034	&	0.043	&	0.078	&	-0.025	\\
	&	(0.0232[7])	&	(0.02 [$^{+0.02}_{-0.04}$])	&		&	(0.034 [$^{+0.004}_{-0.006}$])	&	(0.060 [$^{+0.006}_{-0.008}$])	&	(-0.016 [$^{+0.003}_{-0.003}$])	\\
$4_3\rightarrow 2_1$	&	-0.102	&	-0.066	&	-0.014	&	0.018	&	-0.025	&	-0.02	\\
	&	($<|0.13|$)	&		&		&		&		&		\\
$6_3\rightarrow 4_1$	&	0.141	&	-0.045	&	-0.0551	&	-0.028	&	0.011	&	-0.048	\\
$8_3\rightarrow 6_1$	&	0.102	&	0.07	&	-0.077	&	0.012	&	-0.104	&	-0.122	\\
$10_3\rightarrow 8_1$	&	0.007	&	-0.056	&	-0.002	&	0.004	&	0.029	&	0.141	\\
													
\hline													
\multicolumn{7}{c}{$I_{0_2^+} \rightarrow (I-2)_{\gamma}$ }\\													
\cline{1-7}													
$4_3\rightarrow 2_2$	&	0.003	&	0.052	&	-0.011	&	0.229	&	0.041	&	-0.132	\\
$6_3\rightarrow 4_2$	&	0.078	&	-0.026	&	-0.067	&	-0.139	&	0.017	&	0.122	\\
$8_3\rightarrow 6_2$	&	-0.001	&	-0.102	&	-0.167	&	-0.02	&	-0.767	&	-0.007	\\
$10_3\rightarrow 8_2$	&	0.004	&	0.069	&	-0.248	&	0.062	&	1.395	&	-0.14	\\

\hline \hline
  \label{E2T4}
\end{longtable*}

\begin{longtable*}{p{2.72cm}p{2.72cm}p{2.72cm}p{2.72cm}p{2.72cm}p{2.72cm}c}
  \caption{Reduced $E2$ in-band matrix elements (in eb units) for transitions $I_i \rightarrow (I-1)_f$ in $^{76}$Ge, $^{76,78,80,82}$Se and $^{100}$Mo isotopes. Experimental and associated errors are in parenthesis. The signs in parenthesis behind  the TPSM numbers correspond to the signs of the numerical TPSM eigenvectors. 
  }\\

  \hline\hline
$I_i \rightarrow I_f$	&	$^{70}$Ge	&	$^{76}$Se	&	 $^{78}$Se	&	$^{80}$Se	&	$^{82}$Se	&	$^{100}$Mo	\\
													
\hline\hline
\endfirsthead													
													
\multicolumn{7}{c}{$I_{\gamma} \rightarrow (I-1)_{\gamma}$ }\\													
\cline{1-7}													
$3_1\rightarrow 2_2$	&	0.869	&	0.642	&	1.029	&	0.677	&	0.772	&	0.881	\\
$4_2\rightarrow 3_1$	&	-0.082	&	0.319	&	0.79	&	0.326	&	0.498	&	-0.255	\\
$5_1\rightarrow 4_2$	&	0.76	&	0.482	&	0.956	&	-0.517	&	-0.671	&	0.657	\\
$6_2\rightarrow 5_1$	&	0.508	&	0.212	&	0.46	&	-0.206	&	-0.241	&	0.115	\\
$7_1\rightarrow 6_2$	&	-0.639	&	0.381	&	0.687	&	0.291	&	0.023	&	0.521	\\
$8_2\rightarrow 7_1$	&	0.652	&	0.036	&	0.579	&	0.205	&	-0.137	&	-0.195	\\
$9_1\rightarrow 8_2$	&	0.564	&	-0.076	&	-0.585	&	-0.24	&	0.401	&	-0.358	\\
$10_2\rightarrow 9_1$	&	0.034	&	0.031	&	0.651	&	0.122	&	0.785	&	0.365	\\
						
 \hline \hline

  \label{E2T5}
\end{longtable*}

\begin{longtable*} {p{2.72cm}p{2.72cm}p{2.72cm}p{2.72cm}p{2.72cm}p{2.72cm}c}
  \caption{Reduced $E2$ inter-band matrix elements (in eb units) for transitions $I_i \rightarrow (I-1)_f$ in $^{76}$Ge, $^{76,78,80,82}$Se and $^{100}$Mo isotopes. Experimental and associated errors are in parenthesis. The signs in parenthesis behind  the TPSM numbers correspond to the signs of the numerical TPSM eigenvectors. 
  }\\

  \hline\hline
$I_i \rightarrow I_f$	&	$^{70}$Ge	&	$^{76}$Se	&	 $^{78}$Se	&	$^{80}$Se	&	$^{82}$Se	&	$^{100}$Mo	\\
													
\hline\hline
\endfirsthead													

\multicolumn{7}{c}{$I_{\gamma} \rightarrow (I-1)_{Yrast}$ }\\													
\cline{1-7}													
$3_1\rightarrow 2_1$	&	-0.033	&	-0.08	&	0.261	&	-0.188	&	-0.503	&	0.032	\\
	&		&		&		&		&		&		\\
$5_1\rightarrow 4_1$	&	-0.121	&	-0.119	&	-0.228	&	0.138	&	0.551	&	-0.077	\\
$7_1\rightarrow 6_1$	&	0.531	&	-0.15	&	-0.236	&	-0.11	&	-0.117	&	-0.123	\\
$9_1\rightarrow 8_1$	&	-0.796	&	0.135	&	-0.483	&	-0.285	&	-0.869	&	-0.301	\\
													
\hline													
\multicolumn{7}{c}{$I_{0_2^+} \rightarrow (I-1)_{\gamma }$ }\\													
\cline{1-7}													
$4_3\rightarrow 3_1$	&	0.069	&	0.054	&	-0.019	&	-0.031	&	0.131	&	-0.487	\\
$6_3\rightarrow 5_1$	&	-0.036	&	0.132	&	0.098	&	0.128	&	0.036	&	0.049	\\
$8_3\rightarrow 7_1$	&	-0.229	&	-0.014	&	0.073	&	-0.158	&	0.092	&	0.054	\\
$10_3\rightarrow 9_1$	&	-0.032	&	-0.008	&	0.106	&	-0.013	&	-0.195	&	-0.076	\\

 \hline \hline

  \label{E2T6}
\end{longtable*}

\begin{longtable*}{p{2.72cm}p{2.72cm}p{2.72cm}p{2.72cm}p{2.72cm}p{2.72cm}c}
  \caption{Reduced $E2$ inter-band matrix elements (in eb units) for transitions $I_i \rightarrow (I+1)_f$ in $^{76}$Ge, $^{76,78,80,82}$Se and $^{100}$Mo isotopes. Experimental and associated errors are in parenthesis. The signs in parenthesis behind  the TPSM numbers correspond to the signs of the numerical TPSM eigenvectors. 
  }\\

  \hline\hline
$I_i \rightarrow I_f$	&	$^{70}$Ge	&	$^{76}$Se	&	 $^{78}$Se	&	$^{80}$Se	&	$^{82}$Se	&	$^{100}$Mo	\\
													
\hline\hline
\endfirsthead													
													
\multicolumn{7}{c}{$I_{\gamma} \rightarrow (I+1)_{Yrast}$ }\\													
\cline{1-7}													
$3_1\rightarrow 4_1$	&	0.018	&	-0.38	&	0.017	&	0.0003	&	-0.045	&	0.027	\\
$5_1\rightarrow 6_1$	&	-0.54	&	-0.341	&	-0.363	&	0.381	&	0.618	&	-0.53	\\
$7_1\rightarrow 8_1$	&	-0.474	&	0.353	&	0.264	&	0.19	&	0.077	&	-0.46	\\
$9_1\rightarrow 10_1$	&	-0.834	&	0.308	&	-0.262	&	-0.306	&	-0.421	&	-0.313	\\

\hline													
\multicolumn{7}{c}{$I_{0_2^+} \rightarrow (I+1)_{\gamma}$}\\													
\cline{1-7}													
$2_3\rightarrow 3_1$	&	0.056	&	-0.167	&	0.051	&	-0.062	&	0.072	&	-0.047	\\
$4_3\rightarrow 5_1$	&	0.275	&	-0.023	&	0.153	&	0.043	&	0.032	&	0.156	\\
$6_3\rightarrow 7_1$	&	-0.004	&	0.076	&	0.027	&	-0.159	&	-0.985	&	-0.074	\\
$8_3\rightarrow 9_1$	&	-0.176	&	-0.155	&	0.043	&	0.156	&	0.508	&	-0.092	\\
		
 \hline \hline

  \label{E2T7}
\end{longtable*}
\begin{longtable*}{p{2.52cm}p{2.52cm}p{2.52cm}p{2.52cm}p{2.52cm}p{2.52cm}c}
  \caption{Reduced $E2$ inter-band matrix elements (in eb units) for transitions $I_i \rightarrow (I+2)_f$ in $^{76}$Ge, $^{76,78,80,82}$Se and $^{100}$Mo isotopes. Experimental and associated errors are in parenthesis. The signs in parenthesis behind  the TPSM numbers correspond to the signs of the numerical TPSM eigenvectors. 
  }\\

\hline\hline
$I_i \rightarrow I_f$	&	$^{70}$Ge	&	$^{76}$Se	&	 $^{78}$Se	&	$^{80}$Se	&	$^{82}$Se	&	$^{100}$Mo	\\
													
\hline\hline
\endfirsthead													
													
\multicolumn{7}{c}{$I_{\gamma} \rightarrow (I+2)_{Yrast}$ }\\													
\cline{1-7}													
$2_2\rightarrow 4_1$	&	-0.218	&	0.137	&	-0.041	&	-0.06	&	0.077	&	0.33	\\
	&	(-0.43 [$^{+2}_{-3}$])	&	(0.09 [$^{+0.12}_{-0.09}$])	&		&	( $\pm$0.09 [$^{+0.04}_{-0.05}$])	&	(0.08 [$^{+0.04}_{-0.13}$])	&	(0.77 [$^{+0.13}_{-0.10}$])	\\
$4_2\rightarrow 6_1$	&	0.332	&	0.028	&	0.077	&	0.58	&	0.113	&	0.052	\\
	&	($<|0.9|$)	&		&		&		&	(-0.08 [$^{+0.23}_{-0.72}$])	&		\\
$6_2\rightarrow 8_1$	&	0.364	&	0.031	&	0.328	&	0.311	&	1.033	&	-0.146	\\
$8_2\rightarrow 10_1$	&	0.884	&	-0.038	&	-0.1	&	-0.024	&	0.133	&	0.328	\\
													
\hline													
\multicolumn{7}{c}{$I_{0_2^+} \rightarrow (I+2)_{Yrast}$ }\\													
\cline{1-7}													
$0_2\rightarrow 2_1$	&	0.199	&	0.68	&	0.004	&	0.188	&	-0.082	&	0.479	\\
	&	(0.232 [$^{+4}_{-11}$])	&	(0.47 [$^{+0.11}_{-0.10}$])	&	(0.18 [$^{+0.06}_{-0.06}$])	&	( 0.12 [$^{+0.01}_{-0.01}$])	&	(-0.11 [$^{+0.02}_{-0.01}$])	&	(0.513 [$^{+0.009}_{-0.004}$])	\\
$2_3\rightarrow 4_1$	&	0.43	&	-0.046	&	-0.091	&	0.031	&	0.171	&	0.623	\\
	&	(0.59 [4])	&		&		&		&		&	(0.83 [$^{+0.07}_{-0.04}$])	\\
$4_3\rightarrow 6_1$	&	0.572	&	0.165	&	-0.248	&	-0.061	&	-0.015	&	0.026	\\
$6_3\rightarrow 8_1$	&	0.111	&	0.055	&	-0.022	&	0.047	&	-0.008	&	0.187	\\
$8_3\rightarrow 10_1$	&	-0.014	&	0.065	&	0.09	&	-0.121	&	-0.297	&	0.036	\\
													
\hline													
\multicolumn{7}{c}{$I_{0_2^+} \rightarrow (I+2)_{\gamma}$}\\													
\cline{1-7}													
$0_2\rightarrow 2_2$	&	0.365	&	0.162	&	-0.014	&	-0.066	&	0.029	&	-0.466	\\
	&	(0.291 [$^{+10}_{-8}$])	&	(0.15[$^{+0.08}_{-0.18}$])	&		&	(-0.05 [$^{+0.01}_{-0.05}$])	&	(0.06 [$^{+0.17}_{-0.04}$])	&	(-0.32[$^{+0.03}_{-0.02}$])	\\
$2_3\rightarrow 4_2$	&	0.31	&	-0.028	&	-0.04	&	0.127	&	0.373	&	0.027	\\
	&		&	(0.79 [$^{+0.27}_{-0.13}$])	&		&		&	(0.60 [{$^{+0.07}_{-0.07}$}])	&		\\
$4_3\rightarrow 6_2$	&	-0.309	&	-0.062	&	-0.115	&	-0.107	&	0.066	&	-0.029	\\
$6_3\rightarrow 8_2$	&	0.269	&	-0.166	&	0.012	&	0.14	&	-0.059	&	0.146	\\
$8_3\rightarrow 10_2$	&	0.073	&	-0.126	&	0.004	&	0.127	&	-1.142	&	0.178	\\

 \hline \hline

  \label{E2T8}
\end{longtable*}

\bibliographystyle{apsrev4-2}
\bibliography{E2_paper}

\end{document}